\documentclass[12pt]{article}
\usepackage{bm}
\usepackage[onehalfspacing]{setspace}

\usepackage[final]{changes}
\makeatletter
\colorlet{Changes@Color}{teal}
\makeatother

\usepackage{natbib}
\usepackage{graphicx}
\usepackage{caption}
\usepackage{subcaption}
\usepackage{amsmath,amssymb,amsfonts,mathtools}
\usepackage{color}
\usepackage{relsize}
\usepackage{pgfplots}
\usepackage{enumerate}
\usepackage{dirtytalk}
\pgfplotsset{compat=1.12}

\newtheorem{corollary}{Corollary}
\newtheorem{prop}{Proposition}

\begin{document}

\title{Optimal Verification of (Mis)Information in Networks}

\author{Luca P. Merlino and Nicole Tabasso\thanks{\scriptsize{
Merlino: ECARES, Universit\'{e} libre de Bruxelles, Ave. F.D. Roosevelt 50, CP 114/04, Belgium; email: luca.paolo.merlino@ulb.be. Tabasso: Ca' Foscari University of Venice, Venice, Italy; email: nicole.tabasso@unive.it. Merlino gratefully acknowledges financial support from the FWO (grant G029621N), the National Bank of Belgium and the Belgian Research Funds (FNRS, grant T.0289.26). Tabasso gratefully acknowledges funding from the European Union's Horizon 2020 research and innovation programme under the Marie Skłodowska-Curie grant agreement No. 793769. Part of the work was carried out within the project Rumors on Networks and received funding from the European Union Next-GenerationEU--National Recovery and Resilience Plan (NRRP)-–Mission 4 Component 2, Investment N.1.2 – CUP N.H73C22001340001. The work reflects only the authors’ views and opinions, neither the European Union nor the European Commission can be considered responsible for them.\protect\includegraphics[height=.9\baselineskip]{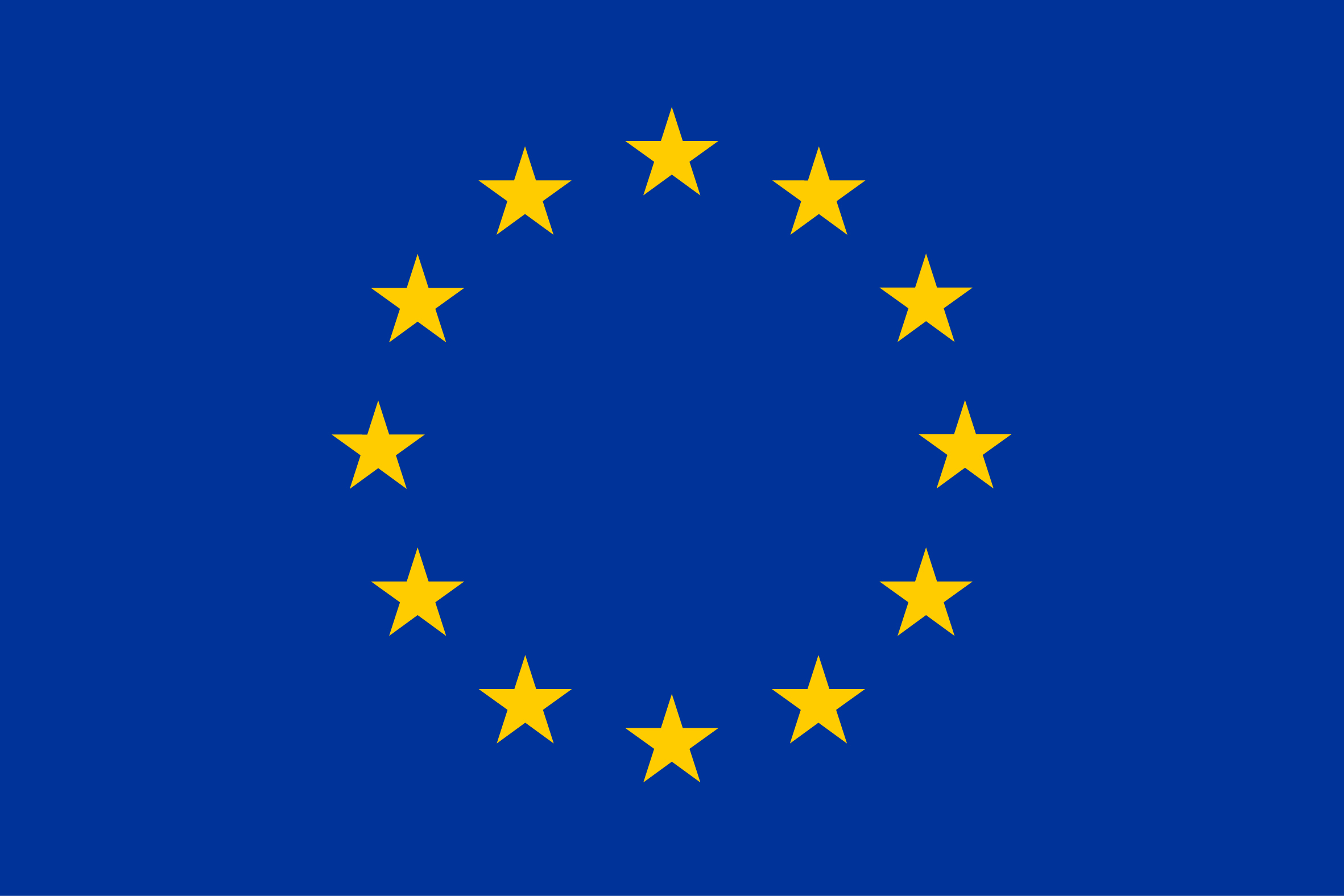}} We thank A. Rusinowska, M. Elliott, S. Sarangi and conference participants at CTN Moscow \& Barcelona, $7^{th}$ \& $8^{th}$ Conferences on Network Science and Economics, ASSET 2023, CES Networks and Games Paris and the Cambridge Janeway Networks Webinar for useful feedback.}}

\date{ }

\maketitle
\vspace{-1em} 
\begin{abstract}
We study the diffusion of a true and a false message (misinformation) when agents are biased and able to verify messages. As a recipient of a false message who verifies it becomes informed of the truth, a higher prevalence of misinformation can increase the prevalence of the truth. We uncover conditions such that this happens and discuss policy implications. Specifically, a planner aiming to maximize the prevalence of the truth should allow misinformation to circulate if: non-verified messages may be ignored, transmission of information is relatively low, and the planner's budget to induce verification is neither too low nor too high. Homophily increases the spread of misinformation, but also facilitates diffusion of truth, and leads to similar results on the effect of verification.
	
    
\end{abstract}

\thispagestyle{empty}
\clearpage
\pagenumbering{arabic} 
\section{Introduction}

The diffusion of misinformation, or \say{fake news}, has received considerable attention in recent years (e.g., \citealp{Allcott,lazer2018}). Yet, such information generally diffuses simultaneously with correct information, and possible interactions are often overlooked in the quest to minimize the diffusion of misinformation. In some cases, the prevalence of the truth may be the socially relevant variable, e.g., when being aware of the truth makes a person more likely to adopt a correct behavior, while being misinformed implies taking the same action as an uninformed agent. For example, being aware that HIV is a sexually transmitted disease makes it more likely for individuals to have protected sexual contacts rather than unprotected ones, leading furthermore to positive externalities. Broadly, this situation occurs naturally whenever the truth requires a specific change in behavior, such as when a new disease is discovered.

The diffusion of information on social networks is a complex matter, and various policies have been suggested to curb the spread of misinformation. For example, policy makers or online social platforms can influence agents' incentives to verify through various channels. These include direct ones, such as raising \textit{information literacy} rates or publishing guides on how to spot fake news, as done by, e.g., \textit{The New York Times} or \textit{Le Monde}, as well as indirect ones, by investing in education in general.

In this paper, we focus on one particular aspect, namely the rate at which agents verify messages they receive. In particular, our main question of interest is the verification rate that a benevolent planner, whose goal is to maximize the proportion of correctly informed agents, would set.\footnote{While in the main model we assume verification rates are exogenous, \added{in Section \ref{microfoundation}} we microfound the problem of choosing verification rates with a model in which agents individually choose whether to acquire the ability to verify information, e.g., through schooling. A planner may impact these rates by subsidizing schooling.}

To the best of our knowledge, we are the first to uncover the conditions under which increased verification reduces both the prevalence of incorrect \textit{and} correct information and crucially, when this is not the case. We find that there are scenarios where some misinformation that could be eradicated may be allowed to circulate as it ``creates truth''. We show that this result arises because misinformation creates awareness of an issue, which chimes with the marketing concept that bad publicity may be useful \citep{berger2010positive}.

We model the diffusion of information using an amended version of the $SIS$ (\textit{Susceptible}-\textit{Infected}-\textit{Susceptible}) framework.\footnote{While initially developed in epidemiology, early examples in economics include, e.g., \cite{lopez2008diffusion} and \cite{jackson2013diffusion}. \cite{Tabasso1} is an early example of a model with multiple infected states.} In this framework, the network is modeled as the number of meetings each agent has per period. On this network, two messages regarding the true state of the world spread through word of mouth, as a result of which there exist two infected states ($I_0$ and $I_1$) as agents may believe either one of them. One message is accurate, while the other is not (which we denote as misinformation), but agents do not know in advance which one is correct. Agents that have not received a message are unaware that there is a pay-off relevant state of the world at all, and receive a utility of zero. \added{Agents who are misinformed also receive a utility of zero, while correctly informed agents obtain a positive payoff, normalized to one.}\footnote{\added{This payoff structure implies that the planner's objective of maximizing the prevalence of truth corresponds to maximizing aggregate welfare, treating misinformed and susceptible agents symmetrically.}}

Agents belong to one of two types, each biased towards believing one of these messages. Importantly, agents who do not verify ignore messages not in line with their bias.\footnote{This assumption captures the concept of \textit{information avoidance} \citep{golman2017information}, which we discuss in detail in Section \ref{model}.} Verification instead is able to reveal the veracity of information. Consequently, irrespective of which message agents receive, if they verify it, they become aware of the true state of the world. Finally, agents only pass on information they believe to their neighbors.

We find that, in steady state, misinformation prevalence is strictly decreasing in verification rates; in fact, high enough verification rates are able to entirely eradicate misinformation. The prevalence of the truth is increasing in verification rates if misinformation is eliminated; but if it survives, truth prevalence is actually increasing in misinformation prevalence. Indeed, as verification of incorrect information reveals the truth, there are some agents who become aware of the truth after receiving misinformation. This is particularly relevant for those agents who, absent verification, would ignore the truth. Thus, an increase in verification rates may either increase or decrease the prevalence of the truth.

For a planner aiming to maximize the truth's prevalence, the optimal policy depends on the available budget and the information prevalence. For either a very low or very high budget, it is optimal to use it all for verification, if possible until all misinformation is eliminated. However, for intermediate levels of the budget, it may be better to induce lower verification rates, which allows misinformation to circulate. \added{This is true only if the baseline prevalence of truth—absent any intervention—is low,} as otherwise there is little benefit in fostering misinformation.

This result implies that a central planner may optimally choose to allow misinformation to circulate, even if they have sufficient resources to eradicate it. This insight challenges the intuition that making it easier to assess the veracity of information must necessarily be beneficial to society. 

We extend the model in several directions. First, we introduce homophily, whereby people of the same type are more likely to interact among themselves. We show that homophily favors the diffusion of misinformation, but also of the truth, and may indeed strengthen our results. If the planner could target verification rates to the group biased against the truth, i.e., the one that determines the survival of misinformation, they may choose to target also the other group to let some misinformation circulate. Additionally, the eradication of misinformation is the optimal policy if misinformation causes too much harm, while optimal verification rates are decreasing in the benefits it might create. Importantly, misinformation might still be allowed to circulate even if believing it confers non-zero costs.

Our model highlights the importance of information loss in the transmission process. Indeed, our results continue to hold as long as there are enough agents who ignore unverified messages against their bias.

We focus on the problem a planner faces when they are able to set verification rates or, alternatively, affect agents' incentives to verify messages through policy. To focus on the strategic decision of this rate, we simplify the diffusion process of information; hence, our approach is complementary to models of strategic diffusion of messages \citep{Bloch-Rumors,kranton2024,bravard23,acemoglu2024model} or Bayesian learning \citep{banerjee1993economics,banerjee2004word,papanastasiou2020fake,mani-etal}. As in \cite{mpt22}, we are interested in the diffusion of messages in the presence of misinformation. However, they assume no information loss during transmission. We discuss the differences in assumptions and their implications in Section \ref{model}, after presenting our model.

Verification in our paper acts as vaccination against a disease, as it inoculates agents against believing misinformation. This relates our work to papers on the strategic decision to protect against a disease \citep{Galeotti-Rogers-AEJ,Goyal-Vigier,talamas20,bizzarri2025homophily}. In contrast to these studies, we examine not how protection affects the harmful state, but how it influences the prevalence of the truth, a positive state. Furthermore, unlike these papers, here protection is not a local public good \citep{KM,KM23}.

Related to our work, \cite{Tabasso1} and \cite{clz24} study the simultaneous diffusion of two types of information. However, in these papers, agents may believe in both simultaneously as they are not contradictory.

The paper proceeds as follows. Section \ref{model} introduces the model. Section \ref{main} presents the results of the diffusion process. Section \ref{planner} solves the planner's problem. Section \ref{extensions} presents some extensions of our model. Section \ref{conclusions} concludes. All proofs are in the Appendix.


\section{The Model}\label{model}

In this section, we formally introduce the elements of the model, derive the differential equations that characterize the diffusion process, and set up the planner's problem. Finally, we discuss the main assumptions of the model.

\bigskip

\noindent \textbf{Information.} Time, indexed by $t$, is continuous. There exist two messages $m\in\{0,1\}$ that diffuse simultaneously among agents. These messages convey information about the state of the world, $\mu\in\{0,1\}$. Without loss of generality, we assume that the true state of the world, unknown to the agents, is $\mu=0$. We refer to $m=0$ as the ``truth'', and to $m=1$ as ``misinformation''.

\bigskip

\textbf{Agents \& Verification.} We consider an infinite population of mass $1$. Agents, indexed by $i$, are in one of three possible states: state $S$ (\textit{Susceptible}—agents who do not hold an opinion and are susceptible to forming one), state $I_0$ (agents believing message $0$) or state $I_1$ (agents believing message $1$).

The population is divided into two groups, indexed by $b \in {0,1}$. A share $x \in [0,1)$ of the population is of type $b=0$, and the remaining (strictly positive) share $1-x$ is of type $b=1$. Before defining what an agent’s type represents, we introduce the notion of message verification. We assume that a fraction $\alpha \in [0,1]$ of the population verifies a message upon receiving it.\footnote{While we take verification as fixed, Section \ref{microfoundation} considers how it can be interpreted as the reduced form of a setting in which verification is an individual choice and the planner influences its cost. Additionally, we assume verification rates are independent of the agent's type and the message received. We discuss the role of this assumption below and relax it in Section \ref{twoalphas}.}

Upon verification, the agent learns the true state of the world with certainty and accepts it.

An agent’s type reflects their \textit{information bias}: they accept an unverified message only if it aligns with their bias, disregarding it otherwise. Thus, after receiving message $m$, an agent of type $b$ believes it if either \textit{(i)} the message aligns with their type ($m = b$), regardless of verification, or \textit{(ii)} the message does not align with their type ($m \neq b$) but has been verified, revealing it to be correct.

Once agents believe a message (i.e., enter either state $I_0$ or $I_1$), they do not update their belief until they die at rate $\delta$, and are replaced by identical agents in state $S$.\footnote{Alternatively, $\delta$ can be interpreted as the rate at which agents forget received information—for example, due to limited memory or delays between receiving information and acting on it. Nonetheless, in many scenarios, $\delta$ is likely very small given the rapid diffusion of information. Our model accommodates all values of $\delta \geq 0$.}

\begin{figure}[ht!]
 \centering
 \includegraphics[width=12cm]{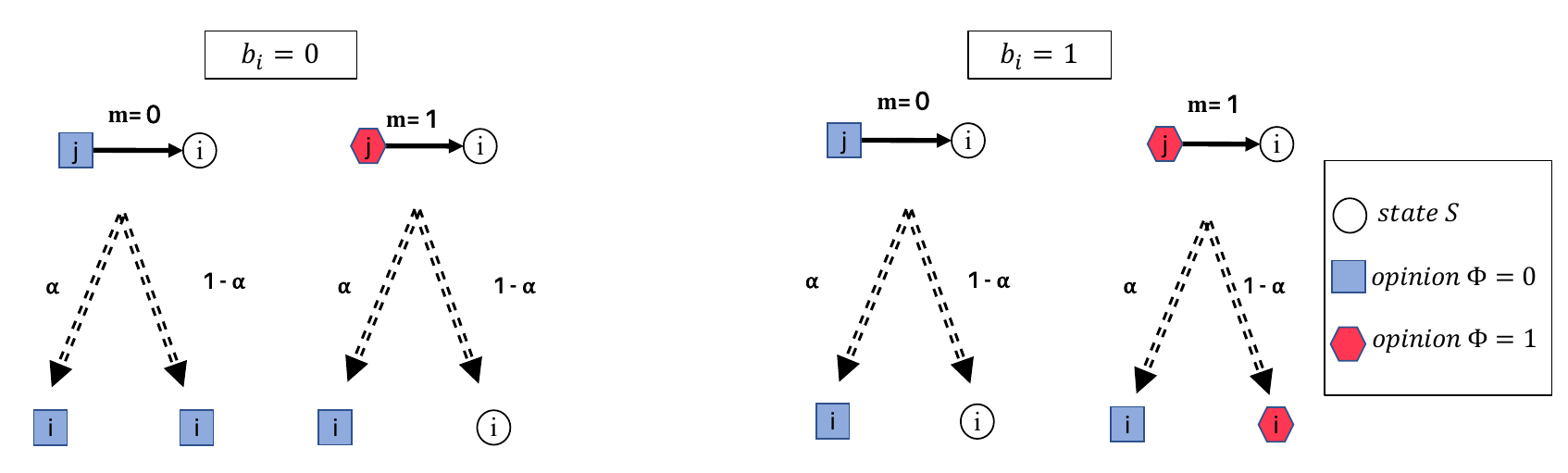}
 \caption{A summary of the potential opinions an agent $i$ may hold, depending on her type, the message received by agent $j$, and verification.}
 \label{fig:opinions}
\end{figure}

To sum up, agents are in state $S$ if they are unaware of both messages, or if they ignore a message they have received. An ``infected'' agent remains in the state (whether $I_0$ or $I_1$) independently of the messages received, unless it transitions back to the susceptible state $S$ at an exogenously determined rate. Once susceptible, the agent may change beliefs again, consistent with the dynamics of the SIS model. Figure \ref{fig:opinions} summarizes which opinion an informed agent holds depending on her type, the message received, and verification.

\bigskip

\noindent \textbf{Network of Diffusion.} A link between agents $i$ and $j$ represents a meeting. These meetings form a communication network, which is independently realized each period. Formally, we model the \textit{mean-field approximation} of the system.

Each agent $i$ has $k$ meetings at $t$, which we refer to as the network density and which is constant over time. We denote by $\nu$ the per-contact transmission rate of message $m$, which is independent of agent type and influenced, for example, by communication technology.\footnote{We assume a degenerate degree distribution, but the model easily accommodates more general ones.}

\bigskip

\noindent \textbf{Information Prevalence.} We denote by $\rho_{b,m,t}^{\alpha}$ the share of type-$b$ agents at time $t$ who believe message $m$ \emph{after verifying it}, for $b\in\{0,1\}$. 
Similarly, $\rho_{b,m,t}^{1-\alpha}$ denotes the share of type-$b$ agents at time $t$ who believe message $m$ \emph{without verifying it}.
\footnote{With a slight abuse of notation, we omit the dependence of the $\rho$'s on the number of meetings $k$, which is identical across agents.}

Because of agents' susceptibility to messages, agents never believe a message they are biased against if they do not verify, nor do they believe misinformation if they verify. As a result, the following equalities hold:
\[
\rho_{0,1,t}^{\alpha}=\rho_{0,1,t}^{1-\alpha}=\rho_{1,0,t}^{1-\alpha}=\rho_{1,1,t}^{\alpha}=0.
\]

We denote the proportion of agents in state $I_0$ and $I_1$ \added{at} time $t$, and therefore overall prevalence of truth ($m=0$) and misinformation ($m=1$) in the population, respectively by
\begin{align}
\rho_{0,t}(\alpha) &= x\Big(\alpha \rho_{0,0,t}^{\alpha}+(1-\alpha)\rho_{0,0,t}^{1-\alpha}\Big) 
+ (1-x)\alpha\rho_{1,0,t}^{\alpha}, \label{theta0}\\ 
\rho_{1,t}(\alpha) &= (1-x)(1-\alpha)\rho_{1,1,t}^{1-\alpha}. \label{theta1}
\end{align}
Equivalently, $\rho_{m,t}$ also denotes the probability a randomly chosen contact of an agent believes message $m\in\{0,1\}$.

We assume that the per contact transmission rate, $\nu$, is sufficiently small that an agent in state $S$ becomes aware of message $m$ at rate $k\nu\rho_{m,t}$ through meeting $k$ neighbors, for $m\in\{0,1\}$.\footnote{This assumption is in line with mean-field analysis, and implies that $\nu$ is sufficiently small that the probability of receiving more than one messages at any given time is negligible.} This framework allows us to model information diffusion as a set of differential equations:
\begin{eqnarray}
\added{\frac{\partial \rho_{0,0,t} ^{\alpha}}{\partial t}} &=& \added{(1-\rho_{0,0,t} ^{\alpha})k\nu[\rho_{0,t} +\rho_{1,t} ] - \rho_{0,0,t} ^{\alpha}\delta,} \label{rho0verify} \\
\added{\frac{\partial \rho_{0,0,t} ^{1-\alpha}}{\partial t}} &=& \added{(1-\rho_{0,0,t} ^{1-\alpha})k\nu\rho_{0,t} - \rho_{0,0,t} ^{1-\alpha}\delta,}\label{rho0nonverify} \\ 
\added{\frac{\partial \rho_{1,0,t} ^{\alpha}}{\partial t}} &=& \added{(1-\rho_{1,0,t} ^{\alpha})k\nu[\rho_{0,t} +\rho_{1,t} ] - \rho_{1,0,t} ^{\alpha}\delta,} \label{rho0group1} \\
\added{\frac{\partial \rho_{1,1,t} ^{1-\alpha}}{\partial t} } &=& \added{(1-\rho_{1,1,t} ^{1-\alpha})k\nu\rho_{1,t} - \rho_{1,1,t} ^{1-\alpha}\delta.} \label{rho1} 
\end{eqnarray}

These differential equations capture the inflow and outflow of agents in each group over time. As an example, take equation \eqref{rho0verify}, which governs the prevalence of truth among \added{the proportion of} verifying type-$0$ agents. The positive term represents the inflow of new believers in message $m=0$: 
only \added{the fraction of} those who have not yet believed message $0$ before time $t$ ($1-\rho_{0,0,t}^{\alpha}$) remain susceptible to believe either of the two messages. Each of these agents meets $k$ neighbors per period, and with probability $\nu$ per contact, they interact. Since a fraction $\rho_{0,t}+\rho_{1,t}$ of the population currently believes either the true message ($I_0$) or the misinformation ($I_1$), their overall exposure rate to information is exactly $k\nu(\rho_{0,t}+\rho_{1,t})$. Because these agents verify, \added{regardless of which message they receive,} they will only believe the true message $0$ upon exposure, as verification rules out believing the false message $1$. 
\added{Note that $\alpha$ does not appear explicitly in this equation since $\rho_{0,0,t}^{\alpha}$ already tracks dynamics within the verifier subpopulation; $\alpha$ shapes the system indirectly through the aggregate prevalence terms $\rho_{0,t}$ and $\rho_{1,t}$ in equations \eqref{theta0}--\eqref{theta1}.} 
The negative term reflects the outflow due to mortality: a fraction $\delta$ of type-$0$ verifiers who currently believe message $0$ exit the population each period. The remaining equations have analogous interpretations.

\bigskip

\noindent \textbf{Steady State.} We focus on the steady state of the system, where equations \eqref{rho0verify}-\eqref{rho1} equal zero. We drop the time subscript $t$ to indicate the steady state value of variables. We define a \textit{Trivial Equilibrium} as a steady state in which both types of information have zero prevalence, and \textit{Non-trivial} otherwise.
We say a steady state $(\rho_{0,0}^{\alpha}, \rho_{0,0}^{1-\alpha}, \rho_{1,0}^{\alpha}, \rho_{1,1}^{1-\alpha})$ of the system (\ref{rho0verify})-(\ref{rho1}) is \textit{globally stable} if for any initial condition $(\rho_{0,0,0}^{\alpha}, \rho_{0,0,0}^{1-\alpha}, \rho_{1,0,0}^{\alpha}, \rho_{1,1,0}^{1-\alpha}) \in (0,1]^4$, the solution satisfies
\[
\lim_{t \to \infty} (\rho_{0,0,t}^{\alpha}, \rho_{0,0,t}^{1-\alpha}, \rho_{1,0,t}^{\alpha}, \rho_{1,1,t}^{1-\alpha}) = (\rho_{0,0}^{\alpha}, \rho_{0,0}^{1-\alpha}, \rho_{1,0}^{\alpha}, \rho_{1,1}^{1-\alpha}).
\]
In words, we say a steady state is stable if the system converges to it after information is seeded in the population.

\bigskip

\noindent \textbf{Social Planner.} We study a social planner’s problem where the policy tool is the population’s verification rate, $\alpha$. The planner has a budget $A$ to induce verification, with a unit cost of one for simplicity.

We assume agents benefit from being correctly informed about the true state, regardless of their type. Being uninformed or misinformed yields the same payoff. The planner’s objective is to maximize the steady-state prevalence of the truth, $\rho_0$. In Section \ref{extensions}, we relax both assumptions.
\bigskip

\noindent \textbf{Discussion of the Main Assumptions.} Before continuing, let us discuss the main assumptions of our model in more detail.

In the model, verification is introduced as a parameter that the social planner can costly influence. This reflects the planner’s ability to affect the costs agents face when deciding how much effort or time to invest in verifying messages, as in \cite{mpt22}. Here, we adopt a reduced-form approach where the planner directly sets the verification rate. Alternatively, the verification rate can be interpreted as the share of the population that is information literate, which the planner may influence through education or digital campaigns aimed at identifying misinformation.\footnote{This interpretation aligns with the assumption that verification rates are determined before message reception; otherwise, verification would vary by message, as in \cite{mpt22}.} Empirical evidence shows that greater education or cognitive sophistication reduces susceptibility to misinformation \citep{bello2022education, pennycook2019lazy}. In Section \ref{extensions}, we show that our reduced-form model can be microfounded by one where individuals decide whether to educate themselves to distinguish truth from misinformation, and the planner subsidizes education to raise verification rates.

In our model, agents verify the information they receive. The results would remain unchanged if, instead, agents verified directly whether they are biased toward the truth after receiving a message. We assume that verification rates are independent of both the type of message and the type of agent receiving it. This reflects the idea they are shaped prior to the diffusion of information—for instance, through education that affects individuals’ ability to identify misinformation (see Section \ref{extensions})—and thus are not specific to a given piece of information. An alternative assumption, following \cite{mpt22}, is that agents verify messages contradicting their priors more frequently than those confirming them. While this would substantially complicate the analysis, it would not alter the mechanism underlying our main result: the steady-state prevalence of truthful information rises with that of misinformation, whereas misinformation declines with higher verification.

A key assumption is that agents’ biases limit their susceptibility to messages, reflecting the tendency to treat contradictory information differently from confirming information and thus filter out negative content \citep{taylor1988illusion}. Essentially, agents exhibit information avoidance by ignoring messages contradicting their opinions while believing those confirming them \citep{golman2017information}. However, this avoidance is not absolute, as agents accept the truth revealed through verification.

Like this paper, \cite{mpt22} study the diffusion of two conflicting pieces of information. The key difference lies in how agents assimilate and transmit information. In \cite{mpt22}, unverified messages lead agents to adopt opinions aligned with their bias; here, agents ignore unverified messages that contradict their bias. These assumptions capture two distinct, well-documented behaviors: ours is grounded in information avoidance \citep[e.g.,][]{taylor1988illusion, golman2017information}, while \cite{mpt22} reflects the observation that exposure to debunking can make agents \textit{more} vocal about their stance \citep[e.g.,][]{zollo2017}.\footnote{Another interpretation of our assumption is that, although only one message is true, multiple types of misinformation exist about the true state, so rejecting a message as false does not necessarily mean knowing the truth. A third interpretation is that agents cannot transmit their opinion directly but only evidence supporting it, which they cannot provide unless they have received or verified it.} In Section \ref{extensions} we present a model where there could be people exhibiting either behavior.

\section{Diffusion of Truth and Misinformation} \label{main}

Denoting the diffusion rate $\lambda$ as $\lambda=\nu/\delta$, a steady state solves
\begin{eqnarray}
\rho_{0,0}^{\alpha} = \rho_{1,0}^{\alpha} &=& \frac{\lambda k[\rho_0+\rho_1]}{1+\lambda k[\rho_0+\rho_1]}, \label{rho0SSv} \\
\rho_{0,0}^{1-\alpha} &=& \frac{\lambda k\rho_0}{1+\lambda k\rho_0}, \label{rho0SSnv}
\\
\rho_{1,1}^{1-\alpha} &=& \frac{\lambda k\rho_1}{1+\lambda k\rho_1}. \label{rho1SS}
\end{eqnarray}
Substituting equations \eqref{rho0SSv}-\eqref{rho1SS} into equations \eqref{theta0} and \eqref{theta1} respectively, the steady states for $\rho_0$ and $\rho_1$ are fixed points of the following expressions:
\begin{eqnarray}
H(\rho_0, \rho_1) &=& \alpha \frac{ \lambda k(\rho_0+\rho_1)}{1+\lambda k(\rho_0+\rho_1)} + x(1-\alpha)\frac{\lambda k\rho_0}{1+\lambda k\rho_0}, \label{theta0SS} \\
G(\rho_1) &=& (1-x)(1-\alpha)\frac{\lambda k\rho_1}{1+\lambda k\rho_1}. \label{Gtheta1SS}
\end{eqnarray}
The first term of equation \eqref{theta0SS} captures the influence of verifying agents: receiving either message leads them to believe the true state is $0$. The second term reflects the additional contribution to truth prevalence from type-$0$ agents who receive $m=0$ without verification. As shown in equation \eqref{Gtheta1SS}, the prevalence of misinformation depends entirely on non-verifying agents. Hence, a steady state of the system is a fixed point of $\rho_0=H(\rho_0, \rho_1)$ conditional on $\rho_1=G(\rho_1)$.

The following proposition characterizes the steady states of the system.
\begin{prop}\label{prop-existence}
   Assume $\alpha>0$. The system of differential equations (\ref{rho0verify})-(\ref{rho1}) admits the following steady states:
\begin{enumerate}[i.]
    \item \textit{(Trivial equilibrium)} The steady state $\rho_0 = \rho_1 = 0$ always exists.
    
    \item \textit{(Truth-only equilibrium)} A steady state with $\rho_0 > 0$ and $\rho_1 = 0$ exists if and only if
    \begin{equation}\label{cond-truth}
    \alpha> \frac{\frac{1}{\lambda k}-x}{1-x}.
    \end{equation}
    In this case, the prevalence of the truth is
    \begin{equation} \label{theta0-norumor}
    \rho_0 = \alpha(1-x)+x-\frac{1}{\lambda k}.
\end{equation}
    \item \textit{(Misinformation equilibrium)} A steady state with $\rho_0 > 0$ and $\rho_1 > 0$ exists if and only if
    \begin{equation}\label{cond-mis}
    \alpha < 1 - \frac{1}{\lambda k(1-x)}.
    \end{equation}
    In this case, $\rho_0$ is given by the strictly positive fixed point of $\rho_0=H(\rho_0, \rho_1)$ given the following value of $\rho_1$:
    \begin{equation}\label{theta1SS}
    \rho_1= (1-\alpha)(1-x) - \frac{1}{\lambda k}.
    \end{equation}
\end{enumerate}
\end{prop}
While the Trivial equilibrium always exists (if nobody is aware of any information, no information can ever be transmitted), the emergence of non-trivial steady states depends on the level of verification. As intuition suggests, high verification levels foster the diffusion of the truth, making the Truth-only equilibrium possible. Conversely, when verification is low, a Misinformation equilibrium arises, where false content spreads among agents biased in favor of misinformation ($b=1$). Both equilibria may coexist (if both condition \eqref{cond-truth} and \eqref{cond-mis} hold ) when $\lambda k>2$, meaning that the diffusion rate is high enough, relative to network density, to sustain positive steady states for either type of information. Another implication of the proposition is that the range of verification rates such that the two equilibria with information diffusion exist is larger if network density $k$ is higher.

An important implication of Proposition \ref{prop-existence} is that a Misinformation-only steady state cannot exist. Since verification rates are positive, verifying agents always uncover the truth; thus, whenever misinformation circulates, so does the truth. We summarize this in the following corollary.
\begin{corollary} \label{coroll:rumorgood}
    If $\alpha>0$, whenever misinformation circulates in steady state ($\rho_1 > 0$), so does the truth ($\rho_0 > 0$).
\end{corollary}
As our goal is to understand how truth and misinformation diffusion respond to verification rates, we next examine the stability of the steady states. 
\begin{prop}\label{prop-stability}
    Assume $\alpha>0$. The system of differential equations (\ref{rho0verify})-(\ref{rho1}) admits a unique globally stable steady state, as follows:
\begin{enumerate}[i.]
    \item if $\alpha\in\left(0 , 1-\frac{1}{\lambda k(1-x)}\right)$, the Misinformation equilibrium is the globally stable steady state.
    
    \item if $\alpha\in\left[1-\frac{1}{\lambda k(1-x)}, \frac{\frac{1}{\lambda k}-x}{1-x}\right]$, the Trivial equilibrium is the globally stable steady state.
    
    \item if $\alpha\in\left(\max\{1-\frac{1}{\lambda k(1-x)},\frac{\frac{1}{\lambda k}-x}{1-x}\}, 1\right]$, the Truth-only equilibrium is the globally stable steady state.
\end{enumerate}
\end{prop}
Proposition \ref{prop-stability} shows that the system always converges to a unique and globally stable steady state, determined by the verification rate $\alpha$. Importantly, whenever both the Misinformation and the Truth-only equilibrium coexist, the Misinformation equilibrium is the uniquely stable one.

Low verification rates are necessary for misinformation to persist, in which case only the Misinformation equilibrium is stable. Conversely, sufficiently high verification rates both eliminate misinformation and foster truth, yielding the Truth-only equilibrium as uniquely stable. What constitutes ``low'' and ``high'' verification rates depends on network and agent characteristics. For instance, if there are more agents biased towards the truth (higher $x$), the parameter range supporting the Misinformation equilibrium shrinks, while that for the Truth-only equilibrium expands. In contrast, increases in either the diffusion rate $\lambda$ or the network density $k$ promote the circulation of both types of information and hence increase the parameter range supporting the Misinformation equilibrium (where both types of information circulate) \added{as globally stable} and shrink that for the Truth-only equilibrium.

At intermediate verification levels, however, neither truth nor misinformation can be sustained, and only the Trivial equilibrium is stable. As Corollary~\ref{coroll:rumorgood} shows, truth always circulates whenever misinformation does. Hence, the range of verification rates that makes the Trivial equilibrium uniquely stable is one in which verification is sufficiently high to suppress misinformation, yet not high enough to sustain truth alone. Such a range exists only if $\lambda k \leq 2$; otherwise, diffusion is strong enough relative to network density that some information inevitably persists.

In summary, equilibrium stability reflects the interplay between message diffusion and the corrective power of verification.

\begin{figure}[ht]
\centering
\includegraphics[scale=0.4]{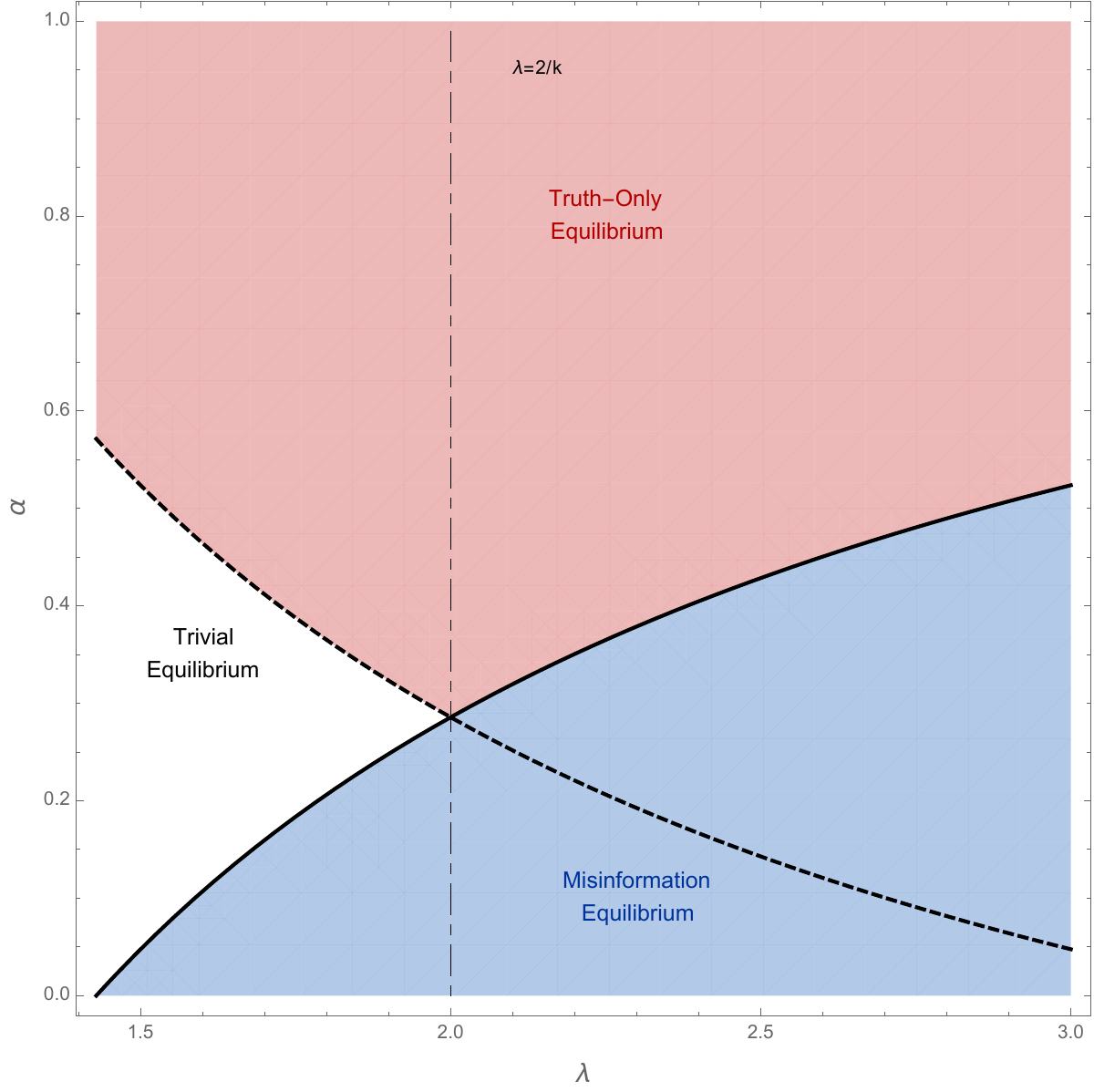}
\caption{Regions of global stability in $(\lambda,\alpha)$ space, for $x=0.3$ and $k=1$. The Misinformation equilibrium (\textcolor{blue}{blue}) is stable for low verification rates; the Truth-Only equilibrium (\textcolor{red}{red}) for high verification rates; and the Trivial equilibrium (gray) for intermediate verification rates when $\lambda k \leq 2$ (left of the vertical dash-dotted line).}
\label{fig-stability}
\end{figure}

\added{Figure \ref{fig-stability} illustrates the three stability regions of Proposition \ref{prop-stability} in $(\lambda, \alpha)$ space, for $x=0.3$ and $k=1$. The solid curve represents the threshold where the verification rate equals $1 - 1/(\lambda k (1-x))$, below which the Misinformation equilibrium is uniquely stable. The dashed curve represents the threshold $(1/(\lambda k) - x)/(1-x)$: the Trivial Equilibrium is uniquely stable if only if the verification rate is between $(1/(\lambda k) - x)/(1-x)$ and $1 - 1/(\lambda k (1-x))$, which is only possible for $\lambda k \leq 2$. Above this threshold, diffusion is strong enough that information of some kind always persists in the network.}

We conclude this section with some comparative statics exercises. Note that these hold irrespective of the stability of the referenced steady states.
\begin{prop} \label{prop-compstats}
Assume $\alpha>0$ and $x\in(0,1)$. Then,
\begin{enumerate}[i.]
    \item In any non-trivial steady state, the prevalence of the truth  ($\rho_0$) increases with the diffusion rate ($\lambda$), the network density ($k$), and the proportion of agents biased in favor of the truth ($x$).

    \item In the Truth-only equilibrium, the prevalence of the truth ($\rho_0$) furthermore increases with the level of verification ($\alpha$).
    
    \item In the Misinformation equilibrium, the prevalence of the truth ($\rho_0$) furthermore increases with the prevalence of misinformation ($\rho_1$). The prevalence of misinformation ($\rho_1$) increases with the diffusion rate ($\lambda$), network density ($k$) and the proportion of agents biased in favor of it ($1-x$) and decreases with the verification rate ($\alpha$).
    \end{enumerate}
\end{prop}

\begin{figure}[ht]
\centering
\includegraphics[scale=0.4]{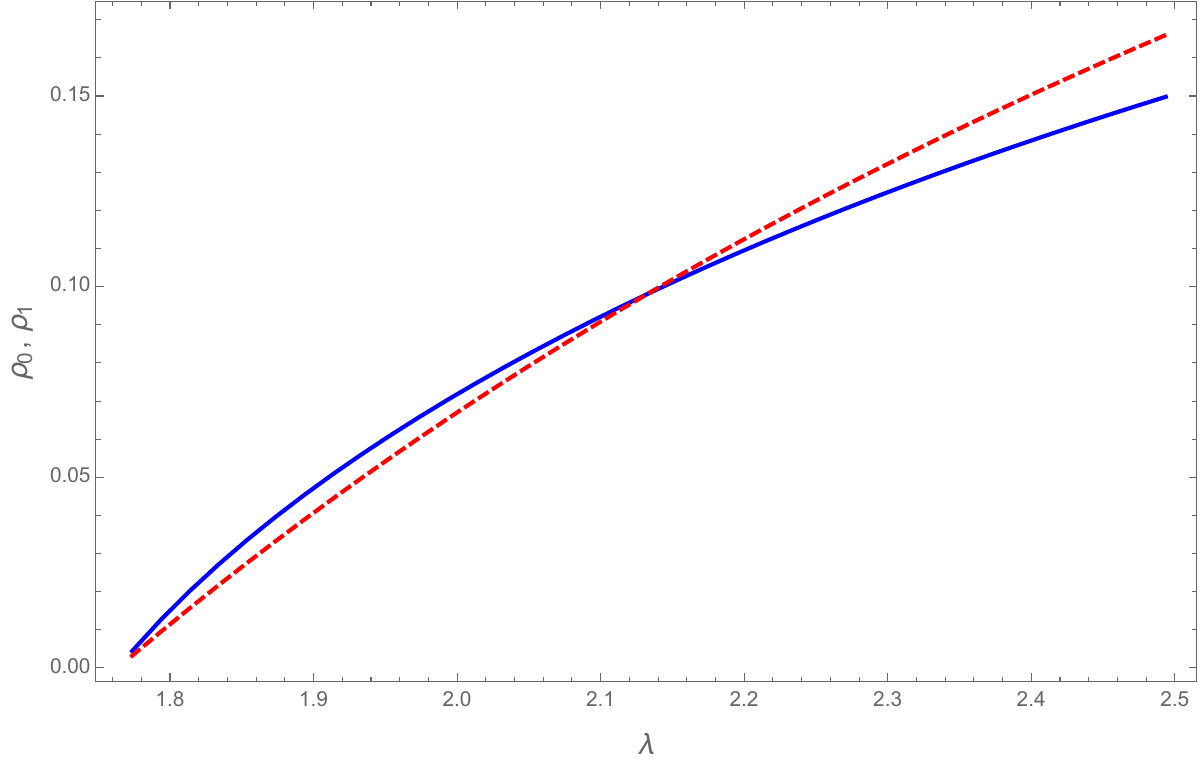}
\caption{\added{Steady state prevalence of truth $\rho_0$ 
(\textcolor{blue}{blue}) and misinformation $\rho_1$ 
(\textcolor{red}{red}, dashed) as functions of the diffusion rate 
$\lambda$, for $\alpha=0.19$, $x=0.3$, and $k=1$.}}
\label{fig-compstats}
\end{figure}

\added{Figure \ref{fig-compstats} illustrates part (iii) of 
Proposition~\ref{prop-compstats}: both $\rho_0$ and $\rho_1$ increase with the diffusion rate $\lambda$ in the Misinformation equilibrium. This is not obvious a priori: one might expect a higher diffusion rate to disproportionately benefit misinformation. Instead, for low values of $\lambda$, the increase in truth prevalence $\rho_0$ exceeds that of misinformation $\rho_1$, so that higher diffusion disproportionately benefits truth. For higher values of $\lambda$, however, the relative slopes reverse and misinformation benefits more. The effect of network density $k$ is analogous, since $\lambda$ and $k$ enter the model multiplicatively.}

\added{More broadly, Proposition~\ref{prop-compstats} (iii) also reveals a counterintuitive effect:} in the Misinformation equilibrium, the spread of misinformation can actually enhance the diffusion of truth---suggesting that, in a sense, ``misinformation creates truth''. \added{This is because, by equation \eqref{theta0SS}, the prevalence of misinformation leads to a higher share of correctly informed agents, for any value of truth prevalence.} Therefore, in environments where verification can overcome bias, allowing some misinformation to circulate may be beneficial. Intuitively, verification promotes the spread of truth and suppresses misinformation. However, since misinformation itself can boost the diffusion of truth, the overall impact of increasing verification is ambiguous---\added{as illustrated in Figure \ref{fig-truthalpha_new} below.} This insight motivates the next section, where we explore the optimal level of verification.

\section{Optimal Verification} \label{planner}

While public discussions often focus on the problem of misinformation, another equally important objective for a planner is the promotion of truth. In many cases, especially in disease prevention, widely sharing accurate information is crucial to encouraging behaviors that improve individual well-being. Knowledge of the truth is often directly linked to the adoption of optimal actions; for example, awareness that HIV is transmitted sexually increases the likelihood of engaging in protected intercourse, both relative to being unaware of the existence of the disease (being in state $S$) and relative to being aware of its existence but believing that it is not sexually transmitted (being in state $I_1$). In such contexts, a benevolent planner may seek to maximize the diffusion of truthful information, for instance by influencing the verification rate $\alpha$ through policies that modify the costs agents incur when verifying messages.

As shown in Section \ref{main}, the diffusion of misinformation significantly affects the diffusion of the truth. The following proposition establishes general results on the planner's optimal use of a budget $A$ when choosing the verification rate $\alpha^\ast$ to maximize the prevalence of the truth, $\rho_0$, under the assumption that verification has a unit cost.\footnote{\added{This objective implicitly assigns equal payoffs to misinformed agents (state $I_1$) and susceptible agents (state $S$), both normalized to zero, while correctly informed agents (state $I_0$) obtain a positive payoff. In this sense, the planner's problem can be interpreted as maximizing a lexicographic utility function that prioritizes the promotion of truth over all other considerations, subject to a feasibility constraint. It is motivated by contexts such as the HIV example above, where the primary policy concern is the direct benefits of correct information, and where the welfare difference between being misinformed and being uninformed may be second-order relative to the benefits of being correctly informed. Section \ref{CostofRumor} discusses how the optimal policy when being misinformed entails different payoffs than being uninformed.}}
\begin{prop} \label{prop-truthalpha}
Assume $\alpha\in\left(0 , 1-\frac{1}{\lambda k(1-x)}\right)\cup \left(\max\{1-\frac{1}{\lambda k}(1-x),\frac{\frac{1}{\lambda k}-x}{1-x}\}, 1\right]$. Let $A$ be the budget available to a planner wishing to maximize the prevalence of the truth in the population in the unique globally stable steady state, and verification rates have a unit cost. Then,
\begin{itemize}
    \item[i)] For all values of the diffusion rate, $\lambda$, the network density, $k$, and the share of agents of group $0$, $x$, there exist values $\underline{A}$ and $\bar{A}$ such that, for all $A<\underline{A}$ and for all $A>\bar{A}$, it is optimal for the planner to use all the budget available to debunk misinformation, i.e., $\alpha^\ast=A$.
    \item[ii)] There exists a value of the diffusion rate $\bar{\lambda}$ such that, for $\lambda<\bar{\lambda}$, there exists a range of $\underline{A}\leq A \leq \bar{A}$ such that it is optimal for the planner not to use all the budget available for debunking, i.e., $\alpha^\ast\in(0,A)$. The value $\bar\lambda$ is decreasing in network density, $k$.
\end{itemize}
\end{prop}
Proposition \ref{prop-truthalpha} establishes that it may be an optimal policy for a planner to \textit{not} fully eradicate misinformation, even if that is possible. At the same time, whenever it is optimal to eradicate misinformation, it is also optimal to spend the entire budget to induce verification. 

\begin{figure}[ht]
\centering
\includegraphics[scale=0.35]{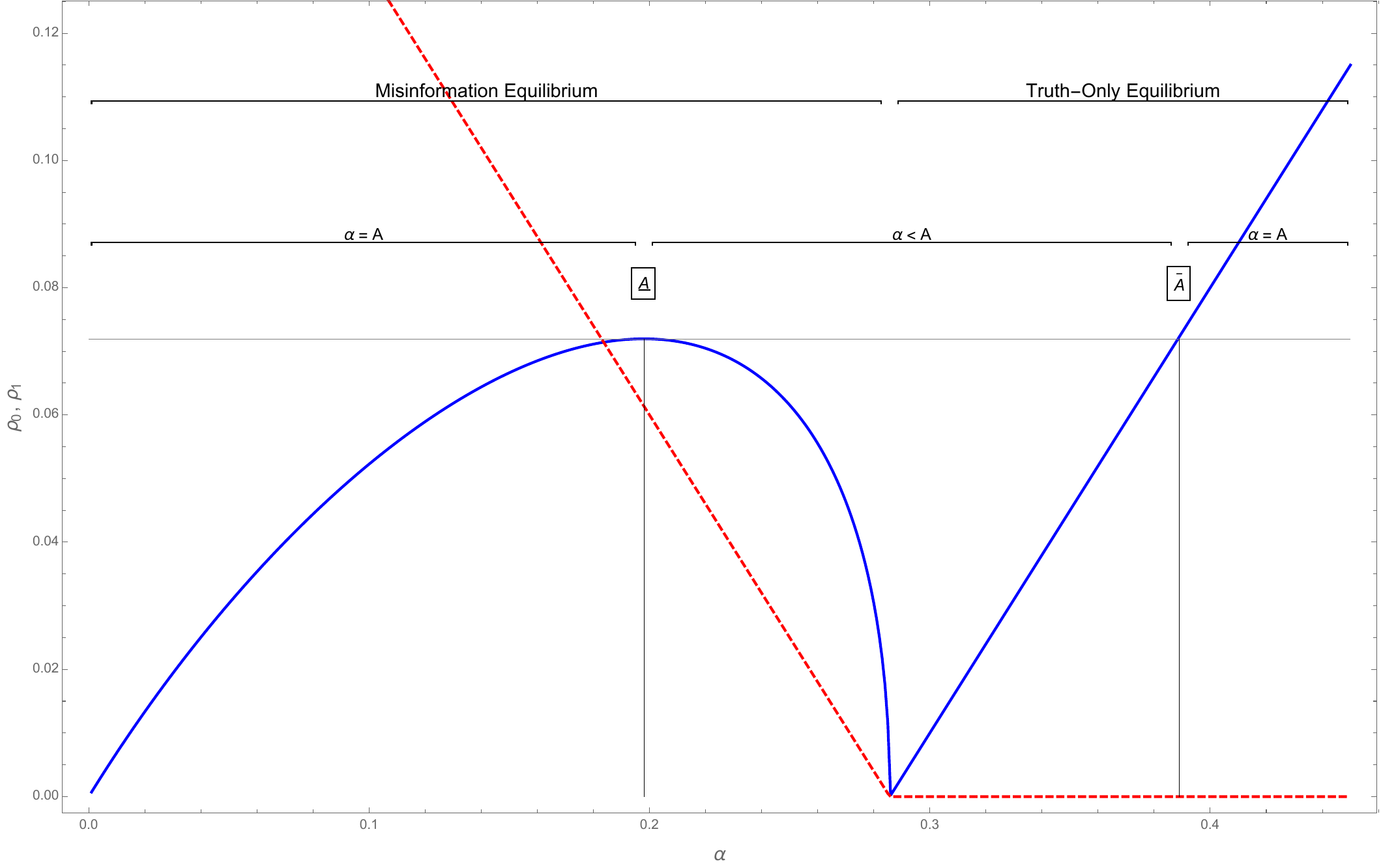}
\caption{Steady state prevalence of truth $\rho_0$ 
(\textcolor{blue}{blue}) and misinformation $\rho_1$ 
(\textcolor{red}{red}, dashed) as functions of $\alpha$, for $\lambda k=2$ and $x=0.3$.}
\label{fig-truthalpha_new}
\end{figure}
 
Intuitively, verification has both a positive and a negative effect on the diffusion of the truth.
On the one hand, it reduces the proportion of agents that ignore truthful messages, which benefits the prevalence of the truth. On the other hand, it reduces the prevalence of misinformation, which causes an indirect effect of fewer messages being transformed into truthful ones. 

Figure \ref{fig-truthalpha_new} highlights how, when verification rates are high enough to eradicate misinformation, only the first, positive, effect is present and truth prevalence is unambiguously increasing in verification rates. If instead verification rates are not sufficiently high to eradicate misinformation, both forces affect the prevalence of the truth. Which effect dominates depends on the verification rate and the initial prevalence of misinformation. When there is little verification or little circulation of misinformation, the indirect effect is very weak; in this case, more verification is good for the prevalence of the truth. Instead, for intermediate levels of verification, as long as there is enough misinformation circulating, the indirect effect dominates the direct effect.


\added{As verification rates depend on the budget available to the planner, the non-monotonic effect is present only for intermediate levels of the budget. These two observations—about how $\rho_0$ responds to $\alpha$, and about how much of the budget to use—are related but distinct. With a low budget $A \leq \underline{A}$, the planner spends it all, as the direct positive effect of verification on truth prevalence when $\alpha\leq \underline{A}$ (remembering that verification carries a unit cost). For intermediate budgets $A \in (\underline{A}, \bar{A})$, the planner optimally spends less than the full budget, as a decrease in verification can increase the truth prevalence. For $A > \bar{A}$, the planner again spends the entire budget: misinformation is eradicated and the marginal return to verification exceeds its unit cost for any value of verification, making it optimal to spend the entire budget on it. Here, $\bar{A}$ is the verification rate in the Truth Only Equilibrium that yields the same truth prevalence as verification rate $\underline{A}$ in the Misinformation Equilibrium.}

Additionally, this effect is present only for relatively low (relative to network density) values of the diffusion rate, $\lambda$. This describes well scenarios where few people talk about an issue, and thus there is only a little communication, so that few people become informed directly.

An example is information about the AIDS-HIV link and sexual transmission of HIV. Our model would predict that in this early phase, with little communication due to a lack of understanding of the disease and stigmatization, an increase in the discourse (even if not correct) would have led to more correctly informed people than a focus on all information being correct. However, once diffusion rates are higher or the network becomes denser, the optimal policy becomes one of minimization of misinformation, and full eradication. Indeed, high diffusion rates or network density imply a high prevalence of the truth, which then lowers the relative importance of the indirect effect of verification. Full eradication, however, requires higher verification rates as diffusion rates and network density increase. For example, while undoubtedly nowadays most people in the world are informed about the link between HIV and AIDS, and most acknowledge that HIV is a sexually transmitted disease, denial of this link is still present and causes significant harm. This is despite a significant drive to educate people and the accessibility of the correct information.\footnote{Take as an extreme example the case where $\delta=0$, such that no informed agent ever forgets or dies. In this case, eradication of misinformation requires that the verification rate is one, i.e., all agents verify.} 

To sum up, the conditions under which it is beneficial for the diffusion of the truth to allow misinformation to circulate are that, \textit{(i)}, verification affects the proportion of agents that are susceptible to information \textit{per se}, \textit{(ii)}, the overall diffusion rate of information is low relative to network density, and, \textit{(iii)}, the budget available is intermediate. We turn now to discuss how these conditions are affected by different specifications of the model.


\section{Discussion} \label{extensions}

\subsection{Homophily} \label{homophily}

So far, we have assumed that agents meet randomly, independent of their type. However, one important feature of real-world networks is that they tend to exhibit significant levels of \textit{homophily}, i.e., the tendency for people to meet those those who are similar to themselves. Following \added{\cite{currarini2009economic}}, we parameterize homophily by a parameter $h\in[0,1]$, so that with probability $h$, agents meet someone of their own type, and with probability $1-h$ they meet others randomly. Hence, setting $h=0$ recovers our benchmark model; conversely, if $h=1$, agents only meet others of their own type.

The total steady state information prevalence of the truth ($\rho_0$) and misinformation ($\rho_1$) respectively are unaffected by the introduction of homophily and remain as presented in equations \eqref{theta0} and \eqref{theta1}:
\begin{eqnarray}
\rho_0 &=& x\big(\alpha \rho_{0,0}^\alpha + (1-\alpha)\rho_{0,0}^{1-\alpha}\big) + (1-x)\big(\alpha \rho_{1,0}^\alpha\big), \nonumber \label{beta_rho0} \\
\rho_1 &=& (1-x)(1-\alpha)\rho_{1,1}^{1-\alpha}. \nonumber\label{beta_rho1}
\end{eqnarray}
Homophily biases the meeting probabilities in favor of agents of the same type and hence these quantities are no longer equivalent to the probabilities of a random agent transmitting the relevant information. The system of differential equations describing the diffusion of messages that results is
\begin{align}
\added{\frac{\partial \rho_{0,0,t}^{\alpha}}{\partial t} }
& =\added{(1-\rho_{0,0,t}^{\alpha}) k\nu\left[[(1-h)x + h][\alpha\rho_{0,0,t}^{\alpha} + (1-\alpha)\rho_{0,0,t}^{1-\alpha}] \right.} \nonumber \\
&\quad \added{ \left. + (1-h)(1-x)[\alpha\rho_{1,0,t}^{\alpha} + (1-\alpha)\rho_{1,1,t}^{1-\alpha}]\right] -\rho_{0,0,t}^{\alpha}\delta, } \label{eq:rho00alpha_h} \\
\added{\frac{\partial \rho_{0,0,t}^{1-\alpha}}{\partial t} }
& \added{=(1-\rho_{0,0,t}^{1-\alpha}) k\nu\left[[(1-h)x + h][\alpha\rho_{0,0,t}^{\alpha} + (1-\alpha)\rho_{0,0,t}^{1-\alpha}] \right.} \nonumber \\
&\quad \added{\left. + (1-h)(1-x)\alpha\rho_{1,0,t}^{\alpha}\right] - \rho_{0,0,t}^{1-\alpha}\delta,} \label{eq:rho001alpha_h} 
\end{align}
\begin{align}
\added{\frac{\partial \rho_{1,0,t}^{\alpha}}{\partial t} }
& \added{=(1-\rho_{1,0,t}^{\alpha}) k\nu\left[(1-h)x[\alpha\rho_{0,0,t}^{\alpha} + (1-\alpha)\rho_{0,0,t}^{1-\alpha}] \right.} \nonumber \\
&\quad \added{\left. + [(1-h)(1-x) + h][\alpha\rho_{1,0,t}^{\alpha} + (1-\alpha)\rho_{1,1,t}^{1-\alpha}]\right]} \nonumber \\
&\quad \added{- \rho_{1,0,t}^{\alpha}\delta,} \label{eq:rho10alpha_h} \\
\added{\frac{\partial \rho_{1,1,t}^{1-\alpha}}{\partial t} }
& \added{=(1-\rho_{1,1,t}^{1-\alpha}) k\nu\left[(1-h)(1-x) + h\rho_{1,1,t}^{1-\alpha}\right] }\nonumber \\
&\quad\added{ - \rho_{1,1,t}^{1-\alpha}\delta.} \label{eq:rho111alpha}
\end{align}
Setting equations \eqref{eq:rho00alpha_h}-\eqref{eq:rho111alpha} to zero, we derive the steady state of the system. 
\added{Since equation \eqref{eq:rho111alpha} is independent of $\rho_{0,0}^{\alpha}$, $\rho_{0,0}^{1-\alpha}$, and $\rho_{1,0}^{\alpha}$, misinformation prevalence $\rho_{1,1}^{1-\alpha}$ can be solved for separately and then treated as a parameter in the remaining equations. We therefore represent the remaining steady state system as $\bm F(\bm\rho, h) = 0$, where $\bm F = (\rho_{0,0}^{\alpha} - F_1,\ \rho_{0,0}^{1-\alpha} - F_2,\ \rho_{1,0}^{\alpha} - F_3)^T$ and $F_1, F_2, F_3$ are the right-hand sides of equations \eqref{eq:rho00alpha_h}--\eqref{eq:rho10alpha_h} evaluated at steady state, for a given value of $\rho_{1,1}^{1-\alpha}$.}

We now show the existence, uniqueness, and stability of the Misinformation Equilibrium in which misinformation spreads, to then turn to its analysis.
\begin{prop}\label{h_existence}
    Assume $\alpha>0$. The system of differential equations \eqref{eq:rho00alpha_h}-\eqref{eq:rho111alpha} admits a steady state with $\rho_0>0$ and $\rho_1>0$ (the Misinformation equilibrium) if and only if
        \begin{equation}
            \alpha < 1 - \frac{1}{\lambda k[(1-h)(1-x)+h]}. \label{eq:mis_h}
        \end{equation}
    In this case, the strictly positive prevalence of $\rho_0$ is given by equation \eqref{beta_rho0}, where $\rho$ is the strictly positive fixed point of $F(\rho,h)=0$ given that
    \begin{equation}
        \rho_{1,1}^{1-\alpha} = 1 - \frac{1}{\lambda k(1-\alpha)[h + (1-h)(1-x)]}. \label{eq:rho1_h}
    \end{equation}
    If the misinformation equilibrium exists, it is the only globally stable equilibrium.
\end{prop}
An implication of Proposition \ref{h_existence} is that homophily is fostering the diffusion of misinformation. Indeed, condition \eqref{eq:mis_h} reveals that the misinformation equilibrium exists for a broader range of verification rates as homophily increases. Additionally, within this equilibrium, the prevalence of the truth is positively impacted by the prevalence of misinformation, for the same reason as in our benchmark model.

\begin{figure}[ht!]
 \centering
 \includegraphics[width=12cm]{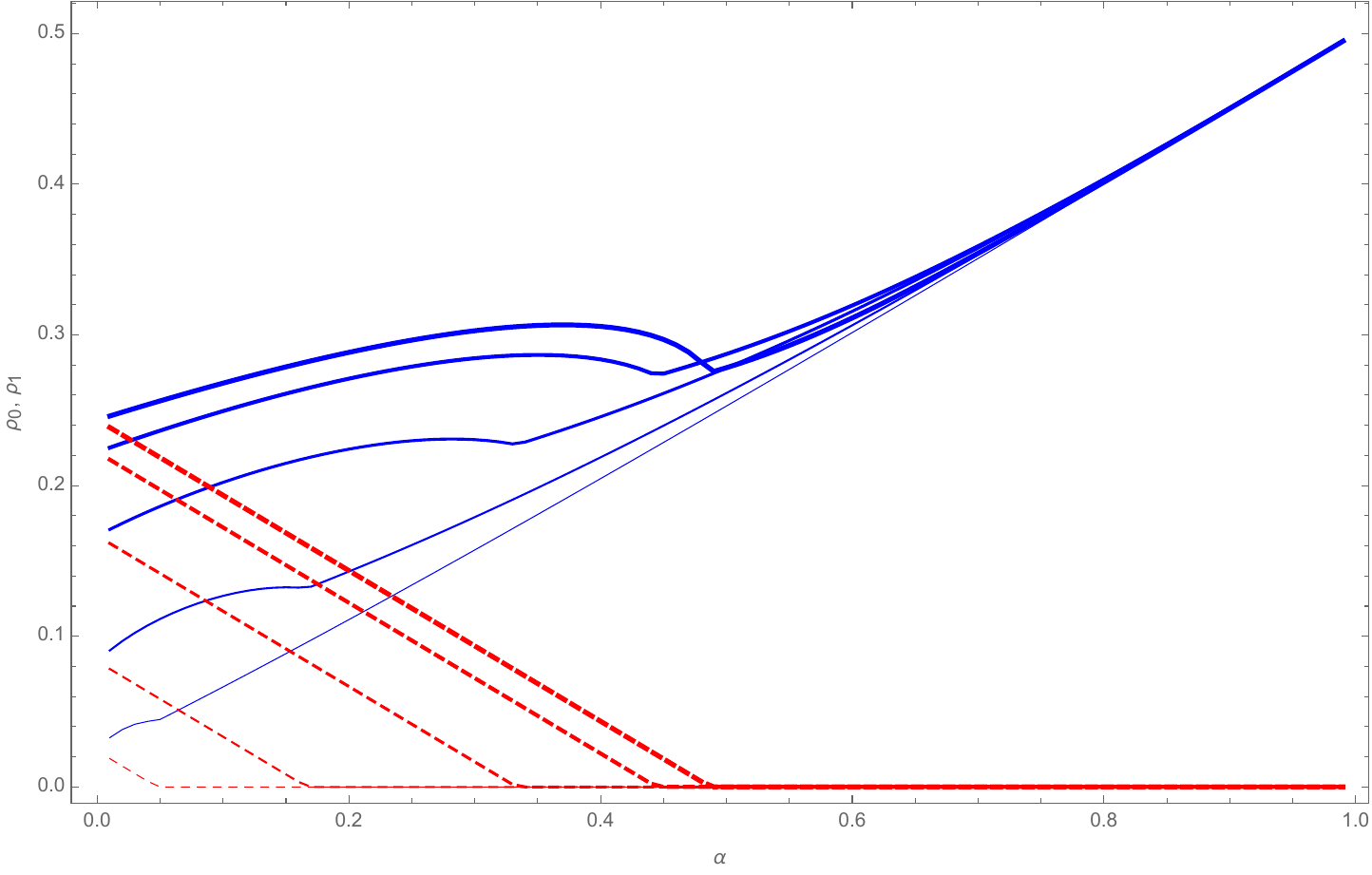}
 \includegraphics[width=12cm]{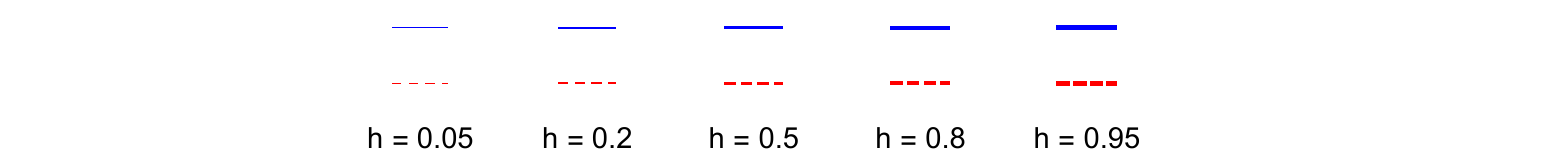}
\caption{\added{Steady state prevalence of truth $\rho_0$ (\textcolor{blue}{blue} lines, solid) and misinformation $\rho_1$ (\textcolor{red}{red} lines, dashed) as functions of $\alpha$, for levels of homophily $h=\{0.05, 0.2, 0.5, 0.8, 0.95\}$, with $\lambda k=2$ and $x=0.55$. Within each color, line thickness increases with $h$, as indicated in the legend. 
}}
 \label{fig:rho0_homophily}
\end{figure}

\added{Figure \ref{fig:rho0_homophily} shows how the prevalence of truth $\rho_0$ and misinformation $\rho_1$ change with verification across different levels of homophily. For low homophily, misinformation is limited and verification unambiguously increases truth prevalence. As homophily rises, misinformation becomes more prevalent and the non-monotonic effect of verification on truth becomes more pronounced: increasing verification may reduce truth prevalence by tightening the indirect channel through which misinformation boosts truth diffusion. For high levels of homophily, misinformation persists even at high verification rates, and the negative effect of verification on truth prevalence is particularly sharp before misinformation is eventually eliminated.}

Figure \ref{fig:rho0_homophily} also highlights that, while homophily benefits the diffusion of misinformation, it may also increase the prevalence of the truth. This positive effect is particularly strong for low values of verification rates, as our next result proves.
\begin{prop}\label{CS_h}
   Consider the steady-state system with homophily parameter $h \in [0,1]$. There exists a level of verification $\bar\alpha$ such that for all $\alpha<\bar\alpha$, the prevalence of the truth is strictly increasing in homophily.
\end{prop}
While it is easily verified that homophily has no effect when everyone verifies ($\alpha=1$),\footnote{At $\alpha=1$, the prevalence of the truth is at its maximal level, $\rho=1-1/(\lambda k)$.} at lower verification rates it supports the circulation of both messages. This occurs because homophily reduces information loss: unverifying agents ignore messages that conflict with their bias, but greater homophily increases encounters with agents of the same group, who are more likely to transmit information that is retained. This effect is especially important when verification is low, making homophily beneficial for the prevalence of the truth. Thus, homophily's positive role challenges the view that it is necessarily harmful by fostering misinformation.

\subsection{Individual Verification Choices} \label{microfoundation}

In the benchmark model, the planner directly chooses the level of verification in the economy. However, this can be seen as a reduced form problem of one in which agents choose verification rates, while the planner affects their costs.

In particular, consider a model in which agents decide verification before the diffusion process begins. The idea is that agents decide at the beginning of their lives whether to educate themselves to allow them to discern truthful from incorrect messages.\footnote{This assumption simplifies the analysis\textemdash see \cite{mpt22} for a model in which agents decide whether to verify messages after they received them. Generalizing the model to allow for verification rates according to the type of message does not affect the main mechanism of our model.} Education bears a cost of $\bar{c}$, and agents receive a flow utility of one if they are aware of the truth and zero otherwise. Agents are infinitely patient, so they only care about the expected proportion of their lives during which they are correctly informed. Each agent is infinitesimal, and hence unable to affect the total proportion of educated agents $\alpha$. Denoting by $\alpha_i$ agent $i$'s verification rate, $i$'s expected lifetime utility is
\begin{equation*}
    U_i(\alpha) =\left\{\begin{array}{cc}
        \frac{\lambda k(\rho_0+\rho_1)}{1+\lambda k(\rho_0+\rho_1)} - \bar{c} & \text{ if } \alpha_i=1, \\
        x\frac{\lambda k\rho_0}{1+\lambda k\rho_0} & \text{ if } \alpha_i=0.
    \end{array}
    \right.
\end{equation*}
For ease of exposition, assume that $\bar{c}$ is so high that, absent the planner's intervention, nobody would get educated and verify.

The planner hence sets a subsidy $s$ (at unit cost) such that
\begin{equation} \label{eq:individual}
    \frac{\lambda k(\rho_0+\rho_1)}{1+\lambda k(\rho_0+\rho_1)} - x\frac{\lambda k\rho_0}{1+\lambda k\rho_0} = \bar{c} - s. 
\end{equation}
The left-hand side of equation \eqref{eq:individual} is always positive, continuous in the verification rate $\alpha$, and smaller than $\bar{c}$ (by assumption). Therefore, the planner can induce any verification rate $\alpha$ by appropriately choosing the subsidy $s$. In particular, the optimal subsidy solves the following problem:
\begin{eqnarray*}
    \max  && \rho_0 \\ 
    s.t.  && \rho_0 = H(\rho_0,\rho_1) \ \ \text{from equation \eqref{theta0SS}} , \\
          && \frac{\lambda k(\rho_0+\rho_1)}{1+\lambda k(\rho_0+\rho_1)} - x\frac{\lambda k\rho_0}{1+\lambda k\rho_0} = \bar{c} - s ,\\
          && \alpha s \leq A , \\
          && \alpha \in (0,1).
\end{eqnarray*}
This problem delivers a solution equivalent to that of the benchmark model. Additionally, if agents differ in their individual costs of education, the planner will subsidize those with lower education costs, and our analysis extends.


\subsection{Agents (Sometimes) Negatively React to Messages}

In our model, agents who receive an unverified message that goes against their bias ignore it\textemdash and hence remain susceptible to future messages. This assumption contrasts to that of \cite{mpt22}, who assume instead that, in such a case, agents become informed of the debate but react negatively to the message they received; hence, they hold an opinion in line with their bias and spread it. As discussed above, the two assumptions capture different aspects of online communication.

The implication is that, while in the present scenario increases in verification rates can have a non-monotonic effect on the prevalence of the truth, in \cite{mpt22} this relationship is always positive.

In reality, people sometimes spread their opinions (such as liking/not liking posts, posting (negative) comments, or posting contradictory information in response to a message they received), and sometimes they simply ignore messages they do not agree with, while re-posting those they find credible.

We now extend our model by introducing a parameter $z$ describing the proportion of the population that ignores unverified messages that go against their bias, while a share $1-z$ of the population converts them into the belief matching their bias.\footnote{As before, $z$ can also be interpreted as the probability with which an agent adopts one behavior vs. the other.} Thus, $z=1$ is the case analyzed in our benchmark model, while $z=0$ corresponds to that of \cite{mpt22}. Setting up the truth and misinformation prevalence as in Section \ref{model} for the two groups and their differential equations, we can derive the steady states of truth and misinformation as the fixed point(s) of the following system:\footnote{These follow straightforwardly the arguments in Section \ref{model}, however the system is now comprised of seven differential equations. The exact equations are available from the authors upon request, as are the explicit solutions referred to below.}
\begin{align}
    \rho_0 = H(\rho_0,\rho_1) &= z\left\{ \alpha\frac{\lambda k(\rho_0+\rho_1)}{1+\lambda k(\rho_0+\rho_1)} + x(1-\alpha)\frac{\lambda k\rho_0}{1+\lambda k\rho_0}  \right\} + \nonumber \label{o0ext}\\
     & \ \ + (1-z)(x+\alpha(1-x))\frac{\lambda k(\rho_0+\rho_1)}{1+\lambda k(\rho_0+\rho_1)}, \\
    \rho_1 = G(\rho_0,\rho_1) &= (1-x)(1-\alpha) \left\{ z\frac{\lambda k\rho_1}{1+\lambda k\rho_1} + (1-z)\frac{\lambda k(\rho_0+\rho_1)}{1+\lambda k(\rho_0+\rho_1)} \right\}. \label{o1ext}
\end{align}
Solving this system shows that the prevalence of the truth is continuous in the share of the population that ignores opposing messages, $z$. Hence, higher verification may decrease the prevalence of the truth also if some agents share their opinions. At the same time, the effect disappears if too many do so.\footnote{The value of $z$ above which truth prevalence is non-monotonic in verification depends on the parameters of the model, but its study does not provide additional insights.}

\subsection{Negative vs. Positive Effects of Misinformation}\label{CostofRumor}

In our benchmark model, we assume that being incorrectly informed and uninformed provides the same utility to agents. This assumption fits well in scenarios such as information about a new disease spreading, where only correctly informed agents can take the correct action. For example, both complete unawareness of the existence of HIV (being in state $S$), as well as believing that it is not a sexually transmitted disease (being in state $I_1$) will have identical incentives for gay men to engage in protected \added{sex. However,} there are situations in which believing misinformation induces actions that entail negative payoffs, or at least, lower payoffs than being entirely uninformed. For example, some home-remedies meant to protect against the COVID-19, such as drinking methanol, ethanol, or bleach, were in fact not only ineffective, but also harmful.

Yet again, believing misinformation might entail positive payoffs, potentially also for the planner. For example, consider an online platform as the planner. Such platforms generally obtain revenues as a function of total engagement. As a result, they may have incentives to maximize the volume of communication (which in our model equals total information prevalence), rather than that of only the truth.

To address these concerns, we extend our model by modifying the social planner's objective function as follows
\begin{equation} \label{eq:rumorcost}
\max \rho_0 + \Phi\rho_1, 
\end{equation}
where $\Phi \in \mathbb{R}$. Hence, our benchmark model corresponds to setting $\Phi=0$, while if $\Phi<0$ or $\Phi>0$ it is costly or beneficial for the planner if agents believe misinformation, respectively. Given the positive steady state prevalence of misinformation it is immediate that
\[
\frac{d 
\added{(\rho_0 + \Phi\rho_1)}}{d\alpha} = \frac{d\rho_0}{d\alpha}-\Phi(1-x),
\]
i.e., \added{the verification incentive has the same dependence on verification, $\alpha$ as in the benchmark problem except for the additional linear term $-\Phi(1-x)$.} 
This implies the following corollary: 

\begin{corollary}
    When the planner maximizes \eqref{eq:rumorcost}, there exists a threshold $\bar{\Phi}$ such that, for all $\Phi<\bar{\Phi}$, the planner uses all the available budget $A$ to eradicate misinformation.
    Otherwise, the optimal verification rate is lower than the available budget $A$.
\end{corollary}

The intuition for the result is as follows. The lower $\Phi<0$ is, the costlier it is to let misinformation circulate. Hence, if these costs are high enough, it is optimal for the planner to use all of the available budget to eradicate misinformation. In particular, when the diffusion rate $\lambda$ is low, the network is sparse, and the budget intermediate, by Proposition \ref{prop-truthalpha}, $\bar{\Phi}$ is negative; in other words, if it is optimal not to use all the budget available for verification when the only objective of the planner is to maximize the truth, the costs associated with misinformation circulating have to be sufficiently large to make it optimal to eradicate it completely.

On the contrary, if $\Phi$ is larger than the threshold $\bar{\Phi}$, i.e., if the benefits associated with misinformation are large enough, it is optimal to let it circulate, either fully or at least to some extent. One noteworthy case is $\Phi=1$, i.e., \added{when the planner maximizes} $\rho_0+\rho_1$. In this scenario, the planner’s objective is simply to maximize the overall prevalence of information, $\rho$, regardless of whether it is true. The total prevalence is given by the fixed point solving
\begin{eqnarray} \label{eq:theta}
    H(\rho,\rho_0,\rho_1) = \alpha \frac{\lambda\rho}{1+\lambda\rho}+x(1-\alpha)\frac{\lambda\rho_0}{1+\lambda\rho_0} + \rho_1.
\end{eqnarray}
The optimal verification rate $\tilde{\alpha}^*$ when $\Phi=1$ relates to the one when only truth is maximized (denoted $\alpha^*$) as follows.
\begin{prop} \label{prop:theta}
Let the planner have budget $A$ available to set verification rates and let their objective be to maximize overall information prevalence. Then, there exists a value of the budget $\tilde{A}>\bar{A}$ such that, for all $A\geq\tilde{A}$, $\tilde{\alpha}^*=\alpha^*=A$. For all $A<\tilde{A}$ instead, the optimal verification rate when maximizing information overall is weakly lower than the one when the planner maximizes only truthful information.
\end{prop}
The lower optimal verification rate under overall prevalence (relative to truth-only maximization) follows from the linear decline of misinformation with verification. More interestingly, for sufficiently high budgets both objectives coincide at the same verification rate. As verification rises, more agents become susceptible to some form of information, raising overall prevalence, which reaches its maximum under full verification.

\subsection{Targeted Verification} \label{twoalphas}

In our model, we assume that verification rates are independent of the type of the agents and the message received. Additionally, message susceptibility of agents is restricted by their type, and this restriction is overcome through the verification of messages. Hence, if no agent was biased towards misinformation, the latter would die out and the truth would achieve its maximum prevalence. This suggests that the verification rate most relevant for our main result—that misinformation may increase the prevalence of truth—is that of agents biased in favor of \added{misinformation}.

For this reason, we now allow the planner to assign different verification rates to the two groups, for example, by targeting information-literacy campaigns or guidelines on identifying misinformation to specific audiences.

In this scenario, the steady state of the truth is the fixed point of 
\begin{small}
\begin{equation*}
    H(\rho_0, \rho_1) = x\left[\alpha_0 \frac{\lambda k(\rho_0+\rho_1)}{1+\lambda k(\rho_0+\rho_1)} + (1-\alpha_0) \frac{\lambda k\rho_0}{1+\lambda k\rho_0}\right] + (1-x)\alpha_1\frac{\lambda k(\rho_0+\rho_1)}{1+\lambda k(\rho_0+\rho_1)}.
\end{equation*}
\end{small}
The prevalence of misinformation instead is given by $\rho_1 = (1-x)(1-\alpha_1)-1/(\lambda k)$ and depends exclusively on verification in group $1$.

Given a budget $A$ and assuming that verification costs are the same in both groups, the planner's problem is the following:
\begin{eqnarray}\label{sp}
\max && \rho_0 \\
\text{s.t.} && \rho_0 = H(\rho_0, \rho_1) \\
&& x \alpha_0+(1-x) \alpha_1 \leq A \\
&& \alpha_0, \alpha_1 \in (0,1). \label{cons}
\end{eqnarray}
In the following, we constrain ourselves to scenarios where $\lambda k>1/(1-x)$, as otherwise misinformation always dies out, independently of verification rates. To focus on the budget allocation across groups, we restrict our attention to budgets $A\leq x$; this implies that for a positive misinformation prevalence, it is always optimal for the planner to use all their budget.\footnote{As this condition implies that all type $0$ agents verify, we do not perceive it as stringent.} Under this condition, we can derive the optimal allocation of resources.
\begin{prop} \label{prop-budget}
The planner's problem described in \eqref{sp}-\eqref{cons} has a unique solution. Furthermore, 
\begin{itemize}
        \item[i)] For all values of the diffusion rate, $\lambda$, and network density $k$, there exist values $A''$ and $A'$, with ${A'}<1-1/(\lambda k(1-x))$, such that for all $A<{A'}$ and for all $A>A''$, it is optimal to debunk \added{misinformation} only in group $1$, i.e., $\alpha_0=0$.
        \item[ii)] For $A\in[A',A'']$, there exist combinations of the budget, $A$, the diffusion rate, $\lambda$, and the network density, $k$, such that it is optimal to debunk \added{misinformation} also in group $0$, i.e., $\alpha_0>0$.
    \end{itemize}
\end{prop}
As Proposition \ref{prop-budget} highlights, the main result of our baseline model, namely that it may be optimal to allow misinformation to circulate, carries over also when the planner can target individual groups to induce verification. Here, this takes the form of diverting resources towards verification in the group biased towards the truth, despite them being insusceptible to misinformation.

\section{Conclusions}\label{conclusions}

In this paper, we model how a true and a false message spread in a population of biased agents who become aware of the veracity of messages they receive if they verify them. We find that the presence of misinformation can foster the diffusion of the truth. This effect can lead to the counterintuitive outcome that increasing verification rates lowers the prevalence of the truth. This happens when: \textit{(i)}, non-verified messages contradicting agents' biases are ignored, \textit{(ii)}, total information prevalence is relatively low, \textit{(iii)}, verification rates are in an intermediate range, and, \textit{(iv)}, believing misinformation does not come at too high costs. Increased homophily may strengthen this effect.

We show that a planner may optimally allow misinformation to persist, even when resources suffice to eliminate it. This challenges the intuition that easier verification is always socially beneficial. To our knowledge, this is the first paper to identify a negative channel through which verification can reduce the prevalence of truth. Whether higher verification ultimately raises or lowers truth in practice remains an open empirical question, but our results stress the need for planners to account for the dynamics of information diffusion when designing verification policies.

In our work, all agents benefit from being aware of the truth, and there are no incentives for agents to diffuse information they themselves do not believe. The inclusion of strategic considerations in information sharing appears a promising avenue for future research.

\pagebreak

\pagebreak
\appendix
\setcounter{equation}{0}
\renewcommand{\theequation}{A-\arabic{equation}}

\section{Proofs} \label{A-alpha}

\noindent \textbf{Proof of Proposition \ref{prop-existence}.} \textbf{Part (i): Trivial equilibrium.} By direct substitution, $(\rho_0, \rho_1) = (0, 0)$ satisfies the steady state conditions for any $\alpha > 0$.

\noindent \textbf{Part (ii): Truth-only equilibrium.} We seek equilibria with $\rho_0 > 0$ and $\rho_1 = 0$. At steady state with $\rho_1 = 0$, the equation for $\rho_0$ becomes
\begin{equation*}
\rho_0 = \alpha + (1-\alpha)x - \frac{1}{\lambda k} = \alpha(1-x) + x - \frac{1}{\lambda k}
\end{equation*}
For $\rho_0 > 0$, we require:
\begin{equation*}
\alpha(1-x) + x - \frac{1}{\lambda k} > 0.
\end{equation*}
Rearranging leads to:
Therefore:
\begin{equation*}
\alpha > \frac{\frac{1}{\lambda k} - x}{1-x}.
\end{equation*}
This establishes condition~\eqref{cond-truth} and formula~\eqref{theta0-norumor}.

\noindent \textbf{Part (iii): Misinformation equilibrium.} We seek equilibria with both $\rho_0 > 0$ and $\rho_1 > 0$. From the steady state condition for $\rho_1$:
\begin{equation*}
\rho_1 = (1-\alpha)(1-x) - \frac{1}{\lambda k}.
\end{equation*}
For $\rho_1 > 0$, we require:
\begin{equation*}
(1-\alpha)(1-x) > \frac{1}{\lambda k},
\end{equation*}
which gives:
\begin{equation*}
\alpha < 1 - \frac{1}{\lambda k(1-x)}.
\end{equation*}
This establishes condition~\eqref{cond-mis} and formula~\eqref{theta1SS}. The existence of a corresponding positive $\rho_0$ in the misinformation equilibrium follows from the coupling in the steady state equations and the requirement that the system be internally consistent. This concludes the proof of Proposition \ref{prop-existence}. \hfill $\blacksquare$

\bigskip

\noindent \textbf{Proof of Proposition \ref{prop-stability}.} The proof follows the arguments of \cite{Jackson-Rogers}. The steady states of $\rho_0$ and $\rho_1$ are fixed points of the system $\rho_0=H(\rho_0,\rho_1)$ as in \eqref{theta0SS} given $\rho_1=G(\rho_1)$ as in \eqref{Gtheta1SS}. Hence, we have:
\\
\noindent \textbf{Part (i)}: By \eqref{Gtheta1SS}, $G(0)=0$, $G(1)<1$, and $G(\rho_1)$ is strictly concave in $\rho_1$. Hence, there exists at most one steady state in which $\rho_1>0$ and it is globally stable if and only if $G'(0)>1$. By simple algebra, this holds if and only if
\begin{equation*}
\alpha < 1 - \frac{1}{\lambda k(1-x)}.
\end{equation*}
Regarding $\rho_0$, $H(\rho_0,\rho_1)$ is also concave in $\rho_0$ with $H(1,\rho_1)<1$ and $H(0,\rho_1)> 0$ with strict inequality if and only if $\rho_1>0$. Concavity of $H(\rho_0,\rho_1)$ ensures that the steady state in which $\rho_0>0$ is globally stable.
\\
\noindent \textbf{Part (iii)}: In the Truth-only equilibrium, $\rho_0>0$ while $\rho_1=0$. Part (i) shows that $\rho_1=0$ exists and is globally stable if and only if
\begin{equation*}
\alpha > 1 - \frac{1}{\lambda k(1-x)}.
\end{equation*}
Regarding $\rho_0$, $H(\rho_0,0)$ is also concave in $\rho_0$ with $H(1,\rho_1)<1$ and $H(0,0)= 0$. Thus, given $\rho_1=0$, there are exactly two steady states for $\rho_0$, one in which $\rho_0=0$ and one in which $\rho_0>0$. Concavity of $H(\rho_0,\rho_1)$ ensures that the steady state in which $\rho_0>0$ is globally stable if such a steady state exists. By proposition \ref{prop-existence}, this yields
\begin{equation*}
\alpha > \frac{\frac{1}{\lambda k} - x}{1-x}.
\end{equation*}
As $\alpha$ needs to satisfy both conditions, part (iii) follows.
\\
\textbf{Part (ii)}: To conclude, by the reasoning above, a Trivial steady state is globally stable if
\begin{equation*}
\alpha \geq 1 - \frac{1}{\lambda k(1-x)}.
\end{equation*}
while
\begin{equation*}
\alpha \leq \frac{\frac{1}{\lambda k} - x}{1-x}.
\end{equation*}
Combining both conditions yields the result, which is possible only if $1 - \frac{1}{\lambda k(1-x)}<\frac{\frac{1}{\lambda k} - x}{1-x}$. This completes the proof of Proposition \ref{prop-stability}. \hfill $\blacksquare$

\bigskip

\noindent \textbf{Proof of Proposition \ref{prop-compstats}.} \textbf{Part (i)}: Independently of the value of $\rho_1$, it is straightforward to establish that $H(\rho_0,\rho_1)$ is increasing in $\lambda, k$, and $x$ for any value of $\rho_0$. As such, increases in these parameters cause an upward shift of $H(\rho_0,\rho_1)$ and therefore increase the steady state $\rho_0>0$.
\\
\textbf{Part (ii)}: Is immediate from equation \eqref{theta0-norumor}. \\
\noindent \textbf{Part (iii)}: The effects of $\lambda, k, 1-x$ and $\alpha$ on $\rho_1>0$ are immediate from equation \eqref{theta1}. The effect of $\rho_1$ on $\rho_0>0$ follows again from the fact that $H(\rho_0,\rho_1)$ is increasing in $\rho_1$ for all values of $\rho_0$. \hfill $\blacksquare$

\bigskip

\noindent \textbf{Proof of Proposition \ref{prop-truthalpha}.} First, note that by equation \eqref{theta0-norumor} the prevalence of the truth is equal to $\rho_0=1-1/(\lambda k)$ if $\alpha=1$. This is the highest value that $\rho_0$ can take. By continuity of $\rho_0$ in $\alpha$, there always exists a value $\bar{A}$, such that it is optimal to set $\alpha=A$ if $A>\bar{A}$.

\noindent Next, assume that the planner's budget is not sufficiently large to fully eradicate misinformation. Hence, the steady state truth prevalence is given by equation \eqref{theta0SS}. By the implicit function theorem, the effect of $\alpha$ on $\rho_0$ is given by
\begin{equation*}
\frac{d \rho_0}{d \alpha} = - \frac{- \frac{\partial H}{\partial \alpha}}{1- \frac{\partial H}{\partial \rho_0}}.
\end{equation*}
As $H(\rho_0, \rho_1)$ is strictly concave in $\rho_0$, we know that at the steady state, $\partial H(\rho_0,\rho_1)/\partial \rho_0<1$. Hence, $d \rho_0/d \alpha>0$ if and only if $\partial H(\rho_0,\rho_1)/\partial \alpha>0$, where
\begin{equation} \label{dHdalpha}
\frac{\partial H(\rho_0,\rho_1)}{\partial \alpha} = \frac{\lambda k(\rho_0+\rho_1)}{1+\lambda k(\rho_0+\rho_1)} - x\frac{\lambda k\rho_0}{1+\lambda k\rho_0} -\alpha(1-x)\frac{\lambda k}{\left[1+\lambda k(\rho_0+\rho_1)\right]^2}.
\end{equation}
As the combination of the first two terms is always positive whenever some information survives, it is obvious from equation \eqref{dHdalpha} that at $\alpha=0$ it is beneficial for the truth to increase verification rates. As $\rho_1$ is strictly decreasing in $\alpha$, for given $\rho_0$, $\partial H(\rho_0,\rho_1)/\partial\alpha$ is strictly decreasing in $\alpha$. Thus, by continuity, setting $\alpha=A$ is optimal for low values of $A$, i.e., for all $A<\underline{A}$.

\noindent Finally, we show when $\partial H(\rho_0,\rho_1)/\partial\alpha$ is negative if $A\in\left[\underline{A},\bar{A}\right]$. As $\partial H(\rho_0,\rho_1)/\partial\alpha$ is decreasing in $\alpha$, we look at its value for the highest possible value of $\alpha$ such that misinformation still survives. In fact, $\rho_1=0$ if $\alpha=1-1/[\lambda k(1-x)]$. The limit of $\partial H(\rho_0,\rho_1)/\partial\alpha$ as $\alpha$ approaches this value is
\begin{equation*}
    \frac{\partial H(\rho_0,\rho_1)}{\partial \alpha} = (1-x) \frac{\lambda k\rho_0}{1+\lambda k\rho_0} - \left[1-x - \frac{1}{\lambda k} \right]\frac{\lambda k}{[1+\lambda k\rho_0]^2},
\end{equation*}
which is negative if
\begin{equation} \label{condtheta}
    \rho_0\left[1+\lambda k\rho_0\right] < 1- \frac{1}{\lambda k(1-x)},
\end{equation}
i.e., for low values of $\rho_0$, and positive for high ones. As $\alpha\rightarrow 1-1/[\lambda k(1-x)]$, we find that $\rho_0 \rightarrow 1 - 2/(\lambda k)$ and therefore condition \eqref{condtheta} is satisfied whenever $\lambda < 1/k[2 + \sqrt{2-1/(1-x)}]$. Continuity of \eqref{dHdalpha} in both $\lambda$ and $\alpha$ then yields the result. \hfill $\blacksquare$
\bigskip

\noindent \textbf{Proof of Proposition \ref{h_existence}.} The result on existence and values of a strictly positive steady state follows directly from the explicit result of $\rho_{1,1}^{1-\alpha}$ provided in the main text. Re-arranging it to solve for the values of $\alpha$ such that it is strictly positive yields the result. Finally, global stability comes from noting that the steady state condition can be written as
\begin{equation*}
    \rho_{1,1}^{1-\alpha} = F_4(\rho_{1,1}^{1-\alpha}, h) = \frac{\lambda k [(1-h)(1-x)+h\rho_{1,1}^{1-\alpha}]}{1+\lambda k [(1-h)(1-x)+h\rho_{1,1}^{1-\alpha}]},
\end{equation*}
i.e., $F_4(\rho_{1,1}^{1-\alpha}, h)$ is strictly increasing and concave in $\rho_{1,1}^{1-\alpha}$ with the standard results of $F_4(0, h)=0$ and $F_4(1, h)<1$, thus the strictly positive steady state, if it exists, is globally stable, concluding the proof of Proposition \ref{h_existence}.
\hfill $\blacksquare$

\bigskip

\noindent \textbf{Proof of Proposition \ref{CS_h}.} Consider the steady-state system $\bm{F}(\bm{\rho}, h) = 0$ with $\bm{F} =(\rho_1-F_1, \rho_2-F_2, \rho_3-F_3)^T$, $\rho_{1}=\rho_{0,0}^\alpha$, $\rho_2=\rho_{0,0}^{1-\alpha}$, $\rho_3=\rho_{1,0}^\alpha$ and $\bm{\rho} = (\rho_1, \rho_2, \rho_3)^T$. For $i=1,2,3$, we get:
\begin{equation*}
F_i = \rho_i - \frac{k\lambda S_i(h)}{1 + k\lambda S_i(h)},
\end{equation*}
where $S_1(h), S_2(h), S_3(h)$ are defined as:
\begin{align*}
 S_1(h) &=h(\alpha\rho_{0,0}^\alpha + (1-\alpha)\rho_{0,0}^{1-\alpha}) \nonumber \\
    &\quad + (1-h)[x(\alpha\rho_{0,0}^\alpha + (1-\alpha)\rho_{0,0}^{1-\alpha}) + (1-x)(\alpha\rho_{1,0}^\alpha + (1-\alpha)C(h))], \\
    S_2(h) &= h(\alpha\rho_{0,0}^\alpha + (1-\alpha)\rho_{0,0}^{1-\alpha}) \nonumber \\
    &\quad + (1-h)[x(\alpha\rho_{0,0}^\alpha + (1-\alpha)\rho_{0,0}^{1-\alpha}) + (1-x)\alpha\rho_{1,0}^\alpha], \\
    S_3(h) &= h(\alpha\rho_{1,0}^\alpha + (1-\alpha)C(h)) \nonumber \\
    &\quad + (1-h)[x(\alpha\rho_{0,0}^\alpha + (1-\alpha)\rho_{0,0}^{1-\alpha}) + (1-x)(\alpha\rho_{1,0}^\alpha + (1-\alpha)C(h))].
\end{align*}
\noindent\textbf{Corner cases of $\alpha=0$.} Note that at $\alpha=0$, $\rho_{0,0}^\alpha$ and $\rho_{1,0}^\alpha$ are zero and in fact do not exist. This implies that we only need to concern ourselves with $S_2(h)$ above, which simplifies to $S_2(h) = \rho_{0,0}^{1-\alpha}[h+(1-h)x]$, and the corresponding steady state can be explicitly solved as
\begin{equation*}
    \rho_{0,0}^{1-\alpha} = 1 - \frac{1}{\lambda k[h+(1-h)x]} \ \ \Rightarrow \rho_0 = x \left[1 - \frac{1}{\lambda k[h+(1-h)x]} \right].
\end{equation*}
It is immediate that this value is strictly increasing in $h$. It remains to show that the system is continuous in $\alpha$ as $\alpha \to 0^+$. In our analysis, we constrain ourselves to values of $\lambda k>\frac{1}{h+(1-h)x}$, i.e., ones where the prevalence of the truth is in fact positive even at $\alpha=0$. 

\noindent \textbf{Continuity at $\alpha\to 0^+$.} If the Jacobian of $F$ is invertible at $\alpha\to 0^+$, our steady states are continuous in $\alpha$. In fact, the Jacobian of our system in the limit $\alpha\to 0^+$ becomes
\[
\lim_{\alpha \to 0^+} \frac{\partial \bm F}{\partial \bm{\rho}} =
\begin{pmatrix}
1 & - \dfrac{\lambda k \bigl[h + (1-h)x\bigr]}{(1 + \lambda k S_1(h))^2} & 0 \\
0 & 1 - \dfrac{\lambda k \bigl[h + (1-h)x\bigr]}{(1 + \lambda k S_2(h))^2} & 0 \\
0 & - \dfrac{\lambda k (1-h)x}{(1 + \lambda k S_3(h))^2} & 1
\end{pmatrix},
\]
with
\begin{align*}
S_1(h) &= [h + (1-h)x] \rho_{0,0}^{1-\alpha} + (1-h)(1-x) C(h),\\
S_2(h) &= [h + (1-h)x] \rho_{0,0}^{1-\alpha},\\
S_3(h) &= (1-h)x \rho_{0,0}^{1-\alpha} + [h + (1-h)(1-x)] C(h).
\end{align*}
which is non-singular as $\det \lim_{\alpha \to 0^+} \frac{\partial \bm F}{\partial \bm{\rho}} \neq 0$ for any value of $\lambda k \neq h + (1-h)x$. Thus, for values of $\alpha<\bar\alpha$, the effect of homophily on the prevalence of the truth is positive. This concludes the proof of Proposition \ref{CS_h}. \hfill $\blacksquare$
\bigskip

\noindent \textbf{Proof of Proposition \ref{prop:theta}.} From the facts that $\rho=\rho_0+\rho_1$ and that $\rho_1$ is decreasing in $\alpha$ linearly, it is immediate that maximization of $\rho$ requires a lower $\alpha$ than maximization of $\rho_0$ whenever $\rho_1>0$. At the same time, note that the fixed point of equation \eqref{eq:theta} can be written as
\begin{eqnarray} \label{Aeq:theta}
\rho = \alpha\frac{\lambda\rho}{1+\lambda\rho} + (1-\alpha)\left[x\frac{\lambda\rho_0}{1+\lambda\rho_0} + (1-x)\frac{\lambda\rho_1}{1+\lambda\rho_1} \right],
\end{eqnarray}
and also note that, if $\alpha=1$, total information prevalence would be identical to the one in the standard $SIS$ model, i.e., $\rho=1-1/(\lambda k)$. Equation \eqref{Aeq:theta} shows that, for all other values of $\alpha$, $\rho$ will be lower than this. By continuity of $\rho$, there must exist a budget $\tilde{A}$ such that above it, it is optimal for the platform to set verification rates equal to the budget. By the fact that $\rho=\rho_0$ when misinformation dies out, this choice is identical for the platform and the planner. Finally, as $\rho>\rho_0$ whenever $\rho_1>0$, it must be the case that $\tilde{A}>\bar{A}$. This concludes the proof of Proposition \ref{prop:theta}. \hfill $\blacksquare$
\bigskip

\noindent \textbf{Proof of Proposition \ref{prop-budget}.} 
Given that it is optimal for the planner to use all their budget whenever $A\leq x$, their problem \eqref{sp} can be rewritten as
\begin{eqnarray}
\max_{\alpha_0\in[0,1]} && \rho_0 \label{sp2}\\
\text{s.t.} && \rho_0^3 (\lambda k)^2+\rho_0^2 B +\rho_0 C+D=0. \notag
\end{eqnarray}
where $B=\lambda k (1+\lambda k-2A\lambda k -2\lambda k x+2\alpha_0\lambda k x)$, $C=\lambda k(1-A-x(1-\alpha_0))(1-A\lambda k-\lambda k x+\alpha_0\lambda k x)$ and $D=A(1-\lambda k+\lambda k A+\lambda k x-\alpha_0\lambda k x)$. Dividing by $(\lambda k)^2$ and after some algebra, the constraint can be rewritten as $f(\alpha_0,\rho_0)=\rho_0^3 (\lambda k)^2+\rho_0^2 b +\rho_0 c+d=0$ with $b=-(2\alpha_1(1-x)+2x-1-1/(\lambda k))/(\lambda k)^2$, $c=-(1-\alpha_1)(1-x)(\alpha_1(1-x)+x-1/(\lambda k))/(\lambda k)^2$ and $d=-\rho_1/(\lambda k)$. Note that $f(\alpha_0,\rho_0)$ is continuous in $\alpha_0$. Furthermore, given that $b$ can be either positive or negative and that $c,d<0$, by Descartes' rule of sign, $f(\alpha_0,\rho_0)=0$ admits at most one positive real solution. If the solution is negative, $\rho_0^\ast=0$. If it is bigger than one, $\rho_0^\ast=1$. This proves existence and uniqueness. 

\noindent Next, we consider the question of whether it is always optimal to prioritize verification in group $1$ above the one in group $0$. First, note that whenever misinformation dies out, the prevalence of the truth becomes $\rho_0=x+(1-x)\alpha_1 -1/(\lambda k)$, strictly increasing in $\alpha_1$ and independent of $\alpha_0$. As in the case of a unique verification rate of the whole population, $\rho_0$ is maximized at $\alpha_1=1$, thus, there always exists a budget $A''$ such that for $A>A''$ it is optimal to invest it entirely in the verification of group $1$ and to set $\alpha_0=0$. Next, the implicit function theorem allows us to study the effect of increases in $\alpha_1$ by determining the sign of
\begin{equation} \label{A-H1}
    \frac{\partial H(\rho_0,\rho_1)}{\partial \alpha_1} = (1-x)\left[ \frac{\lambda k \rho_0}{1+\lambda k\rho_0} - A \frac{\lambda k}{\left[1+\lambda k(\rho_0+\rho_1)\right]^2}\right].
\end{equation}
It is straightforward to show that for given $\rho_0$, equation \eqref{A-H1} is strictly increasing in $\lambda$ and $k$ and decreasing in $A$ and $\alpha_1$. Furthermore, it is negative at $\rho_0=0$, positive at $\rho_0=1$, and strictly increasing in $\rho_0$. At $A=0$, the effect of increasing $\alpha_1$ on the prevalence is positive, strictly so whenever $\rho_0>0$. By continuity of equation \eqref{A-H1} in $A$, we can then always find a value $A'$ such that it is optimal for the planner to set $\alpha_0=0$ if $A<A'$, with one caveat: If $\lambda k\leq 1/(x+(1-x)\alpha_1)$, i.e., the truth only survives if misinformation does, any increase in $\alpha_1$ must be such that misinformation continues to survive. In fact, as $A=1-x-1/(\lambda k)$ is necessary to eradicate misinformation, setting $\alpha_0=0$ is always optimal whenever $A<A'\leq1-x-1/(\lambda k)$.

\noindent Finally, consider the limit of equation \eqref{A-H1} as $\alpha_1\rightarrow 1-1/(\lambda k(1-x))$. In this case $\rho_1\rightarrow 0$ and $\rho_0 \rightarrow 1-2/(\lambda k)$. At these values, equation \eqref{A-H1} shows that truth prevalence increases as $\alpha_1$ is reduced if
\begin{equation*}
-3 +\lambda k +\frac{2}{\lambda k} < A,
\end{equation*}
which means, whenever
\begin{equation}
    \lambda k \in \left(\frac{3+A-[(3+A)^2-8]^{1/2}}{2} , \frac{3+A+[(3+A)^2-8]^{1/2}}{2}\right).
\end{equation}
Due to continuity of equation \eqref{A-H1} in $\alpha_1$, the result follows. This concludes the proof of Proposition \ref{prop-budget}. \hfill $\blacksquare$

\end{document}